\documentclass[12pt]{article}

\usepackage{amssymb}
\usepackage{amsmath}
\usepackage{amscd}
\usepackage{latexsym}

\usepackage{graphicx}

\newcommand{\be}{\begin{equation}}
\newcommand{\ee}{\end{equation}}
\newcommand{\Dlt}{\Delta}
\newcommand{\dlt}{\delta}
\newcommand{\prt}{\partial}
\newcommand{\br}{{\bf r}}

\newcommand{\bt}{\beta}
\newcommand{\vp}{\varphi}
\newcommand{\ep}{\varepsilon}
\newcommand{\al}{\alpha}
\newcommand{\ra}{\rightarrow}
\newcommand{\sgm}{\sigma}

\newcommand{\gm}{\gamma}
\newcommand{\om}{\omega}

\newcommand{\Gm}{\Gamma}

\newcommand{\lbd}{\lambda}

\newcommand{\cH}{{\cal H}}

\newcommand{\cA}{{\cal A}}

\newcommand{\rgl}{\rangle}
\newcommand{\lgl}{\langle}

\begin{document}

\begin{center}

{\Large{\bf Equilibration and thermalization in finite quantum
systems } \\ [5mm]
V.I. Yukalov} \\ [3mm]
{\it Bogolubov Laboratory of Theoretical Physics, \\
 Joint Institute for Nuclear Research, Dubna 141980, Russia \\
      and \\
  National Institute of Optics and Photonics, \\
University of S\~ao Paulo, S\~ao Carlos, Brazil}

\end{center}

\vskip 2cm

\begin{abstract}

Experiments with trapped atomic gases have opened novel possibilities
for studying the evolution of nonequilibrium finite quantum systems,
which revived the necessity of reconsidering and developing the theory
of such processes. This review analyzes the basic approaches to
describing the phenomena of equilibration, thermalization, and
decoherence in finite quantum systems. Isolated, nonisolated, and
quasi-isolated quantum systems are considered. The relations between
equilibration, decoherence, and the existence of time arrow are emphasized.
The possibility for the occurrence of rare events, preventing complete
equilibration, are mentioned.

\end{abstract}

\vskip 3cm


\newpage

\begin{center}

{\Large{\bf Contents}}

\end{center}

\vskip 3mm

{\bf 1. Trapped cold atoms}

\vskip 3mm

   1.1  Introduction

   1.2  Experiments with trapped atoms

   1.3  Numerical simulations with models

\vskip 3mm

{\bf 2. Thermalization versus equilibration}

\vskip 3mm

   2.1  Quantum statistical ensemble

   2.2  Algebra of local observables

   2.3  System statistical state

   2.4  Equilibration of quantum systems

   2.5  Thermalization of quantum systems

\vskip 3mm

{\bf 3. Evolution of statistical states}

\vskip 3mm

   3.1  Representative statistical ensembles

   3.2  Nonequilibrium representative ensembles

   3.3  Evolution of observables quantities

   3.4  Equations in matrix representation

   3.5  Pure quantum states

\vskip 3mm

{\bf 4. Isolated quantum systems}

\vskip 3mm

   4.1  Absence of absolute equilibrium

   4.2  Possibility of equilibrium on average

   4.3  Eigenstate thermalization on average

   4.4  Density of quantum states

   4.5  Statistics of level spacing

\vskip 3mm

{\bf 5. Characteristic time scales}

\vskip 3mm

   5.1  Interaction time

   5.2  Local-equilibrium time

   5.3  Heterophase time

   5.4  Relaxation time

   5.5  Recurrence time

\vskip 3mm

{\bf 6. Nonisolated quantum systems}

\vskip 3mm

   6.1  Quantum system in a bath

   6.2  Distance between statistical operators

   6.3  Decoherence induced by environment

\vskip 3mm

{\bf 7. Quasi-isolated quantum systems}

\vskip 3mm

   7.1  Absence of complete isolation

   7.2  Properties of quasi-isolated systems

   7.3  Equilibration of quasi-isolated systems

   7.4  Examples of decoherence factors

   7.5  Occurrence of rare events

\vskip 3mm

{\bf 8. Equilibration under nondestructive measurements}

\vskip 3mm

   8.1  Definition of nondestructive measurements

   8.2  Temporal dependence of observables

   8.3  Specification of measurement procedure

   8.4  Decoherence under nondestructive measurements

   8.5  Existence of time arrow

\vskip 3mm

{\bf 9. Summary}

\vskip 3mm

\newpage

\section{Trapped cold atoms}

\subsection{Introduction}

In recent years, there has been a remarkable progress in experimental
studies of cold trapped atoms, which has been accompanied by intensive
theoretical investigations. Numerous literature can be found in the
books [1-4] and review articles [5-19]. Experiments with trapped atoms
possess several important features making such experiments one of the
most convenient tools for studying the properties of many-body systems.
First, trapped atomic systems allow for a high degree of tunability,
which makes it possible to vary their parameters in a rather wide range.
Second, the variation of system parameters can be realized very quickly,
thus, putting a system in a nonequilibrium initial state and studying
the system evolution. Third, the diluteness of the cold atomic gases
and exceptionally low temperatures result in rather long timescales
of dynamical effects (typically of the order of milliseconds or longer)
allowing for a very precise time resolution of nonequilibrium processes.
Finally, trapped atomic systems can be well isolated from environment.
Therefore, it is possible to study the quantum dynamics of finite
systems under different levels of isolation and to analyze the
influence of the latter on the system equilibration, decoherence, and
thermalization.

This review is devoted to exactly the latter problems. In the present
section, the most interesting experiments and computer simulations,
considering the dynamic equilibration effects of nonequilibrium trapped
atoms, are surveyed. While the following sections are devoted to the
general theoretical description of such processes. In order to avoid
ambiguities, the review tries to use accurate mathematical formulations.

Throughout the paper, the system of units is employed, where the Planck
and Bolztmann constants are set to unity ($\hbar = 1$, $k_B = 1$).

\subsection{Experiments with trapped atoms}

In the experiment of Greiner et al. [20] $^{87}$Rb atoms were loaded into
a three-dimensional optical lattice. Varying the ratio $U/J$ of the
one-site interactions $U$ to the hopping parameter $J$, the transition
between the superfluid Bose-condensed phase and the incoherent Mott
insulator phase was studied. When atoms were initially in the Mott
insulating phase with the ratio $U/J > 5.8 z_0$, with $z_0$ being the
number of nearest neighbors, and then this ratio was quickly reduced
to $U/J < 5.8 z_0$, corresponding to the superfluid phase, the coherence
was restored during the timescale of $1/J$. In the opposite process
of quench from the Bose-condensed superfluid to the Mott insulator
state, the equilibration was accompanied by collapses and revivals of
the coherent fraction on the shorter time scale of $1/U$ [21]. The
equilibration, after crossing the superfluid-insulator phase transition
line, evolved through the inhomogeneous state with alternating regions
of incoherent Mott insulator phases and coherent superfluid phases [20].

The dynamics of pair condensate formation in a strongly interacting
Fermi gas of $^6$Li, close to Feshbach resonance, was studied by
Zwierlein et al. [22]. The relaxation time, after a rapid magnetic
field ramp across the Feshbach resonance, was of the order of
$100/E_F$. Here $E_F \approx \omega_0 (3 N)^{1/3}$ is the gas Fermi
surface energy, $\omega_0$ is the geometric mean of the trapping
frequencies, and $N$ is the total number of atoms in the trap. After
a rapid ramp, molecular Bose condensate co-existed  with unpaired
Fermi atoms.

Trapped quasi-one-dimensional arrays of Bose gases, each containing
from 40 to 250 $^{87}$Rb atoms, were prepared in an out-of-equilibrium
state by Kinoshita et al. [23]. They observed that the system of two
initially separated Bose-Einstein condensates oscillates, as in a
Newton cradle, but does not noticeably equilibrate even after
thousands of collisions.

Spontaneous symmetry breaking in a quenched spinor Bose-Einstein
condensate was studied by Sadler et al. in a quasi-two-dimensional
trap [24]. Nearly pure Bose condensate of spinor $^{87}$Rb was
prepared in the unmagnetized $m_z = 0$ state at a high quadratic
Zeeman shift. By rapidly reducing the magnitude of the applied
magnetic field, the gas was quenched to low quadratic Zeeman shift,
obtaining conditions that favored the ferromagnetic phase.
Spontaneous symmetry breaking occurred through a heterogeneous stage,
where small ferromagnetic domains of variable size and orientation
were intermixed with unmagnetized regions. This could be regarded
as the phase separation of magnetized and unmagnetized states in the
process of nonequilibrium phase transition. Spin-vortex defects were 
also observed with high confidence in about one-third of all images
containing significant ferromagnetism. Some images indicated as many
as four vortices. All observed vortices were singly quantized, with
no apparent preferred direction or circulation. The thermalization
time was an order longer than the local equilibrium time.

Nonequilibrium coherence dynamics of quasi-one-dimensional Bose
gas of $^{87}$Rb was considered by Hofferberth et al. [25]. The
initial state consisted of two completely separated Bose
condensates loaded into a double well, where they started
interacting and developing the common coherent system. The final
state was a thermalized Bose-condensed gas with a temperature
slightly higher than the initial temperature of separated Bose
condensates. The thermalization time, depending on the system
parameters, was between 6 and 9 ms.

The spontaneous formation of vortices in the process of
nonequilibrium Bose-Einstein condensation and the relation of this
phenomenon to the Kibble-Zurek mechanism [26,27] was investigated
by Weiler et al. [28].

In the experiment of Hung et al. [29], a $^{133}$Cs quantum Bose
gas was trapped in a two-dimensional optical lattice. The lattice
depth was quickly ramped to induce the superfluid to Mott-insulator
phase transition. The global equilibration involved mass transport
and was found to be much slower than the microscopic tunneling
time scale $1/J$.

By using the high spatial resolution in-situ fluorescence imaging,
Bakr et al. [30] and Sherson et al. [31] demonstrated that the
superfluid to Mott insulator phase transition can be studied with
single-site and single-particle resolved detection.

\subsection{Numerical simulations with models}

Numerical simulations for different models play the role of
computer experiments, providing essential information on the
relaxation processes in nonequilibrium finite quantum systems.
In the present subsection, some of such simulations are briefly
described. The detailed analysis of terminology and correct
mathematical definitions will be given in the following sections.
Many of simulations were done for lattice models. Such models are
also of interest for cold atom physics, since periodic atomic
systems are now easily created  by means of optical lattices.

Probably, the first computer simulation of nonequilibrium dynamics
for a finite {\it classical} system was done by Fermi et al. [32] who
considered a one-dimensional chain of anharmonic oscillators and
discovered that it does not equilibrate. But that was a classical
system.

A quantum two-dimensional anharmonic oscillator, corresponding to
the Henon-Heiles model [33], was studied by Feingold et al. [34].
They observed that the mean location and dispersion of the wave
packet, first, tend to quasi-equilibrium values, but later display
large fluctuations.

Jensen and Shankar [35] studied relaxation of a finite spin-half
quantum chain in a magnetic field in two situations, when the model
was integrable or not. They attempted to find out if there would be
difference in the relaxation of the integrable  and nonintegrable
cases. It turned out that both integrable and nonintegrable quantum
systems, with as few as seven degrees of freedom, could exhibit
thermalization in finite time to the states well described by
microcanonical ensemble, provided that the initial state was not
an eigenstate of any considered quantities. Of course, the thermalized
state was not an equilibrium state in the strict sense, but rather a 
quasi-equilibrium state, since the overall behavior of a finite quantum 
system should be quasi-periodic in time. But that was a quasi-equilibrium 
state with a long lifetime corresponding to recurrence time. An 
important observation was that the equilibration of integrable as well 
as nonintegrable systems was very similar to each other.

Rigol et al. [36] and Cassidy et al. [37] simulated the relaxation
of one-dimensional lattice hard-core bosons. This is an integrable
model that equilibrates to a representative Gibbs ensemble [38,39],
differing from the standard ensembles, similarly to the integrable
Luttinger model, as has been shown by Casalilla [40]. A rigorous
definition of the representative statistical ensembles will be
given in the following sections. Generally, the class of
representative ensembles includes the standard microcanonical,
canonical, and grand canonical ensembles. To distinguish between
them, the representative ensembles that are different from the
latter can be called nonstandard.

Comparing the behavior of integrable and nonintegrable systems
for one-dimensional [41-43] and two-dimensional [44] lattices of
hard-core bosons, Rigol et al. [41-44] concluded that integrable
systems relax to a nonstandard representative Gibbs ensemble,
while the nonintegrable systems relax to a standard microcanonical
ensemble. Similar behavior occurs for one-dimensional lattices
of fermions [45].

Kollah et al. [46] investigated one- and two-dimensional optical
lattices described by the boson Hubbard model. They studied the
time evolution following a quench from the superfluid to the
Mott insulator phase. The quench was realized by a sudden change
of the on-site interaction $U$ under a fixed hopping parameter $J$.
The superfluid-insulator line, in the case of unit filling, is
characterized by the critical value $u_c$ of the ratio
$u \equiv U/(J z_0)$, where $z_0$ is the number of nearest neighbors.
For one-dimensional latices, $u_c = 1.8$ and for two-dimensional
lattices, $u_c = 4.2$ (see review [18]). The quench from the
superfluid to insulator phase corresponds to the change of $u$
from $u < u_c$ to the values $u > u_c$. For large values of the
final interaction strength $u \gg u_c$, the system reaches a
quasisteady state that is very different from the standard thermal
equilibrium and retains memory of the initial state. As the final
$u$ decreases, the nature of the steady state changes, and in the
region where $U$ is comparable to $J$, that is, $u \sim 1/z_0$, the
steady state is well approximated by a standard thermally
equilibrium Gibbs ensemble. After the quench, the system exhibits
a number of oscillations with a period $2 \pi / U$, which relax to
a quasisteady state on a time scale $1/J$. Similar results for a
two-dimensional lattice of bosons were found by Natu et al. [47].

The existence of two distinct equilibration regimes seems
surprising, given the nonintegrability of the boson Hubbard model
for any finite values of $U$ and $J$. This tells us that the
equilibration regime and the quasisteady state resulting in this
process depend mainly not on the integrability of the system but
rather on the system parameters. Both, integrable as well as
nonintegrable quantum systems, depending on their parameters,
can relax either to a standard statistical ensemble or to a
nonstandard one, any of such ensembles being just particular cases
of the class of representative ensembles.

The results, similar to the boson Hubbard model were found for a
one-dimensional fermion Hubbard model by Manmana et al. [48]. The
ground-state phase diagram of this model, at half filling, is well
known. For weak on-site interactions $U$, lower than the critical
value, given by $U_c/J = 2$, that is, $u_c = 1$, the system 
consists of a Luttinger liquid, while for $U > U_c$, it is a 
charge-density-wave insulator. By a sudden change of the interaction 
strength it is possible to switch the initial state between metallic 
and insulating phases. The relaxation goes through an oscillatory 
stage, with the oscillations of period $2 \pi / U$, and tends to a 
steady state on a time scale $1/J$. The reached quasistationary 
state is not thermal in the usual sense, since it depends on the 
initial conditions. The time evolutions, starting from different 
initial states, can be distinguished from each other, so that neither
relaxation to one and the same equilibrium state nor the usual
thermalization occur. The resulting quaisisteady state could be
described by a representative ensemble with fixed expectation
values of the powers of the Hamiltonian $<H^n>$. The choice of
these additional conditions can be justified by the fact that,
for a closed system, the set of all powers of the Hamiltonian
constitute an infinite number of integrals of motion.

The considered one-dimensional fermion Hubbard model [48], with
nearest-neighbor interactions, is integrable. The authors have
also studied the effect of adding to the model
next-nearest-neighbor interactions, which makes it nonintegrable.
It has been found that, independently of the model integrability
or criticality, the relaxation process does not change much and
the system always relaxes to a nonthermal quasistationary state
that can be described by a representative ensemble with additional
conditions of fixed $<H^n>$. When different initial states have
the same energy and are sufficiently close to each other, the
observable quantities relax to the same value, i.e., the memory
of the initial state in that case is lost after the relaxation.
However, the general energy distribution is defined by a
representative ensemble retaining the memory of the initial
states.

A chain of two-level systems interacting through the van der
Waals forces, with a strength $U$, was considered by Lesanovsky
et al. [49]. The transitions between the energy levels were due
to the Rabi term, with a Rabi frequency $\Omega$. For weak
interactions $U/\Omega \ll 1$, no equilibration was noticed. But
for sufficiently strong interactions $U/\Omega > 2$, the system
reached a steady state at the time of order $50/\Omega$. The
resulting steady state could be well characterized by a grand
canonical ensemble. Thermalization also occurs for a finite spin
system with randomly distributed initial states [50].

Integrable as well as nonintegrable finite quantum systems, it
seems, both can equilibrate. Integrability does not seem to play
a crucial role on the structure of the quasistationary state.
This is despite the fact that integrable and nonintegrable
quantum systems display different level-spacing statistics and
differently react to external perturbations [51]. Though
integrable systems can equilibrate, but their main difference
from nonintegrable systems can be in much longer equilibration
times. This behavior is contrary to integrable classical finite
systems that do not equilibrate at all. Nonintegrable classical
systems can equilibrate if they are chaotic [52,53].

The influence of the vicinity of critical points on the
nonequilibrium dynamics of closed quantum systems was
investigated by Polkovnikov et al. [54-56] who emphasized the
universality of this dynamics in gapless systems near continuous
phase transitions.

\section{Thermalization versus equilibration}

\subsection{Quantum statistical ensemble}

The term {\it statistical ensemble} is constantly employed in studies
of many-body systems. It is therefore important to recall its correct
mathematical definition.

A quantum system is characterized by a Hilbert space of microstates
$\mathcal{F}$. The probability operator measure is given on
$\mathcal{F}$ by a statistical operator $\hat{\rho}(t)$ parameterized
by the time variable $t \geq 0$. Hence $\hat{\rho}(t)$ is a positive
operator normalized to one,
$$
{\rm Tr} \hat\rho(t) = 1 \;  ,
$$
where the trace is over $\mathcal{F}$. By definition, a
{\it quantum statistical ensemble} is the pair
$\{\mathcal{F}, \hat{\rho}(t)\}$.

\subsection{Algebra of local observables}

On the Hilbert space $\mathcal{F}$, one defines self-adjoint
operators $\hat{A}$ that are called the operators of local
observables, provided their expectation values
\be
\label{1}
\lgl \hat A(t) \rgl \equiv {\rm Tr}\hat\rho(t) \hat A
\ee
correspond to the measurable quantities. The set of all such
operators forms the algebra of local observables
$\cA \equiv \{\hat A \}$.

\subsection{System statistical state}

The set of all expectation values for the operators of local
observables is the {\it statistical state}
\be
\label{2}
 \lgl {\cal A}(t) \rgl \equiv \left \{
\lgl \hat A(t) \rgl \right \} \;  .
\ee
An important role in dynamical problems is played by the {\it initial
statistical state}
\be
\label{3}
\cA_0 \equiv \lgl \cA(0)\rgl = \left \{
\lgl \hat A(0) \rgl \right \} \;  ,
\ee
which is the set of the expectation values for the operators of local
observables,
$$
\lgl \hat A(0) \rgl = {\rm Tr} \hat\rho(0) \hat A \;  ,
$$
at the initial moment of time $t = 0$.

\subsection{Equilibration of quantum systems}

In a review, devoted to equilibration and thermalization, it is
crucial to follow precise definitions of these processes. In
literature, one can meet different descriptions of these phenomena.
Below, the definitions are given that will be used in the present
review.

{\it A statistical system, characterized by a statistical ensemble
$\{\mathcal{F}, \hat{\rho}(t)\}$, equilibrates from an initial
statistical state $\mathcal{A}_0$ if and only if for any
$\hat{A} \in \mathcal{A}$ there exists a limit
\be
\label{4}
 \lim_{t\ra\infty} \lgl \hat A(t) \rgl =
{\rm Tr} \hat\rho^*(\cA_0) \hat A \; ,
\ee
generally, depending on the initial state $\cA_0$}.

If the limit (\ref{4}) exists, one says that the statistical system
is in an {\it equilibrium state} with a statistical ensemble
$\{\mathcal{F}, \hat{\rho}^*(\mathcal{A}_0)\}$.

This is a general definition of equilibration as such. One also
considers another form of equilibration, the equilibration on
average. This will be treated in the following sections. It is
worth stressing that the above definition of equilibration
requires that the limits (\ref{4}) would exist for all operators from
the algebra of local observables. When such limits exist only for
some observables, but are absent for others, it is not, strictly
speaking, equilibration.

Also, when a statistical state appears to be time-independent
for a finite interval of time, but not for the actual limit
$t \ra \infty$, this is termed {\it quasi-equilibration}.

\subsection{Thermalization of quantum systems}

Thermalization is a more restrictive process, when not merely
limit (\ref{4}) exists, but the memory of initial conditions is, at
least partially, lost.

{\it A statistical system, characterized by a statistical ensemble
$\{\mathcal{F}, \hat{\rho}(t)\}$, thermalizes from an initial state
$\mathcal{A}_0$ if and only if there exists a dense set
$\mathcal{B} = \{\mathcal{A}_0\}$ of initial states, including the
given initial state $\mathcal{A}_0$, such that for any
$\hat{A} \in \mathcal{A}$ the limit
\be
\label{5}
 \lim_{t\ra\infty} \lgl \hat A(t) \rgl =
{\rm Tr} \hat\rho^* \hat A
\ee
does not depend on $\mathcal{A}_0 \in \mathcal{B}$}.

The set $\mathcal{B}$ is termed the {\it attraction basin} of the
stationary state (\ref{5}).

When the thermalization occurs from any available initial state,
that is, when the attraction basin is the whole set of all
admissible initial states, this is called the {\it global thermalization}.

\section{Evolution of statistical states}

\subsection{Representative statistical ensembles}

To correctly describe the system equilibration, it is necessary to
be accurate in classifying the limiting stationary states. Each
statistical system is characterized by a statistical ensemble. It
is assumed that the statistical ensemble is defined so that to
uniquely represent the considered statistical system. This implies
that the definition of the ensemble must take into account all
conditions and constraints that correctly represent the system.
This principal point was emphasized by Gibbs [38,39], Tolman [57]
and ter Haar [58]. The term {\it representative ensemble} was
introduced by Tolman [57].

The practical realization for constructing a representative ensemble
can be illustrated for an equilibrium system, following the basic
Gibbs idea. Let a Hilbert space of microstates $\mathcal{F}$ be given.
To define a statistical ensemble $\{\mathcal{F}, \hat{\rho}\}$, one needs
to specify the operator probability measure, that is, the statistical
operator $\hat{\rho}$. In constructing the latter, it is necessary to
take into account those constraints that uniquely define the system.
Of course, the statistical operator is to be normalized as
\be
\label{6}
 {\rm Tr} \hat\rho = 1 \; .
\ee
An important quantity is the internal energy given by the average
of the energy operator $\hat{H}$, called Hamiltonian,
\be
\label{7}
 {\rm Tr} \hat\rho \hat H = E \;  .
\ee
In addition, there can be other constraints defining the expectation
values of some {\it constraint operators} $C_i$, with $i = 1, 2, \ldots$,
\be
\label{8}
{\rm Tr} \hat\rho \hat C_i = C_i \; .
\ee

The statistical operator of an equilibrium state is defined as the
minimizer of the {\it information functional}
\be
\label{9}
I[\hat\rho ] = {\rm Tr} \hat\rho \ln\hat\rho +
\lbd_0 \left ( {\rm Tr}\hat\rho - 1 \right ) 
+ \bt \left ( {\rm Tr}\hat\rho \hat H  - E \right )  +
\bt \sum_i \lbd_i \left ( {\rm Tr}\hat\rho \hat C_i - C_i \right ) \;  ,
\ee
in which $\lambda_0, \lambda_i$, and $\beta$ are Lagrange multipliers.
This is equivalent to the maximization of the Shannon entropy under
conditions (\ref{6})-(\ref{8}). The result is the statistical operator
\be
\label{10}
\hat\rho = \frac{1}{Z} \; e^{-\bt H} \;   ,
\ee
with the normalization factor being the partition function
$$
Z \equiv {\rm Tr} e^{-\bt H} \;  ,
$$
and with the {\it grand Hamiltonian}
\be
\label{11}
 H = \hat H + \sum_i \lbd_i \hat C_i \; .
\ee

As is clear, when the sole constraint is the normalization condition (\ref{6}),
then the statistical operator reduces to the microcanonical form
$$
 \hat\rho = \frac{1}{Z} \qquad
\left ( Z = {\rm Tr} \hat 1 \right ) \; .
$$
If only conditions (\ref{6}) and (\ref{7}) are accepted, one gets the Gibbs 
canonical ensemble. If one also defines the average number of atoms, one gets 
the Gibbs grand canonical ensemble. The general situation, with any number
of additional constraints, is also termed the grand Gibbs ensemble [59,60].

The general form of the grand ensemble, with the statistical operator (\ref{10}),
was introduced by Gibbs [38,39]. One also calls this the generalized or
conditional Gibbs ensemble [61]. But one should not forget that this
ensemble was advanced by Gibbs. When deciding on a particular type of a
Gibbs ensemble, one has to choose such that would be representative for the
considered statistical system. General mathematical properties and
applications of representative statistical ensembles have been analyzed in
Refs. [13, 62-64].

\subsection{Nonequilibrium representative ensembles}

When a quantum system equilibrates, it tends to one of the equilibrium
representative statistical ensembles. A nonequilibrium state is
characterized by a nonequilibrium ensemble $\{\mathcal{F}, \hat{\rho}(t)\}$.
The temporal evolution of the statistical operator is governed by a
unitary evolution operator $\hat{U}(t)$, such that
\be
\label{12}
\hat\rho(t) = \hat U(t) \hat\rho(0) \hat U^+(t) \;  .
\ee
The evolution is generated by the grand Hamiltonian (\ref{11}), and the
evolution operator satisfies the Schr\"{o}dinder equation
\be
\label{13}
i\; \frac{d}{dt} \; \hat U(t) = H\hat U(t) \;  .
\ee
This means that the grand Hamiltonian is the evolution generator.

If one assumes that the evolution is generated by the energy
Hamiltonian $\hat{H}$, then the constraint operators $C_i$ in the
grand Hamiltonian (\ref{11}), defining an equilibrium state, are to be the
integrals of motion, commuting with $\hat{H}$. But if the evolution
is generated by the grand Hamiltonian (\ref{11}) itself, the constraint
operators do not need to be the integrals of motion [13,62-64].

In particular, the constraint operators can fix the required initial
conditions for observable quantities in the form
\be
\label{14}
 C_i = \lgl \hat C_i \rgl =
{\rm Tr}\hat\rho(0) \hat C_i \; .
\ee
These initial conditions then define the solutions to the evolution
equations for the studied observables. Generally, the steady states,
resulting from sudden changes of the Hamiltonian, are not thermal,
in the sense that they retain information on initial conditions [65,66].
As examples of the cases, where the memory of initial conditions is
retained in quasi-equilibrium states, it is possible to recall the
hard-core bosons in lattices [36,37,41-44], Luttinger liquids [40],
Hubbard-type models of bosons and fermions [45-48,67,68], free bosonic
models [69], and one-dimensional fermionic systems with Hall states [70].
Other examples are given by the quasistationary states that are the
solutions to the evolution equations describing the relaxation of
average spins for different strongly nonequilibrium spin systems
(nuclei, electrons, molecules, clusters), whose dynamics essentially
depends on initial conditions, such as the initial spin polarization
[71-80].

It is possible to conclude that closed quantum nonequilibrium systems
equilibrate to quasi-equilibrium states characterized by representative
statistical ensembles. In the majority of cases the resulting
representative ensemble retains the memory of initial states, regardless
of whether a model is integrable or not.

\subsection{Evolution of observable quantities}

What one needs and measures in experiments is the statistical state,
that is, the averages of operators of observables. Therefore,
equilibration has to do with the temporal behavior of such averages.
An average of an operator $\hat{A}$, because of the evolution law
(\ref{12}), can be written in two ways:
\be
\label{15}
 \lgl \hat A(t) \rgl = {\rm Tr}\hat \rho(t)\hat A =
{\rm Tr}\hat\rho \hat A(t) \; ,
\ee
in which
$$
\hat\rho \equiv \hat\rho(0) \; , \qquad
\hat A \equiv \hat A(0) \; ,
$$
and the operator time evolution is given by the Heisenberg form
\be
\label{16}
 \hat A(t) = \hat U^+(t) \hat A(0) \hat U(t) \; .
\ee
The expectation value (\ref{15}) is real, as a consequence of the
operator of an observable being self-adjoint.

The Heisenberg equation of motion for the operator (\ref{16})
reads as
\be
\label{17}
 i \; \frac{d}{dt}\; \hat A(t) = i \hat U^+(t) \;
\frac{\prt\hat A}{\prt t} \; \hat U(t) + \left [
\hat A(t) , \; H(t) \right ] \;  ,
\ee
where
\be
\label{18}
 H(t) \equiv \hat U^+(t) H \hat U(t) \;  .
\ee
The Hamiltonian, being the evolution generator, varies with time only
through an explicit dependence on time, if any,
\be
\label{19}
 i \; \frac{d}{dt}\; H(t) = i \hat U^+(t) \;
\frac{\prt H}{\prt t} \; \hat U(t) \;  .
\ee
If the considered operator does not explicitly depend on time, then
its average evolves in time according to the law
\be
\label{20}
i\; \frac{d}{dt} \lgl \hat A(t) \rgl = \left \lgl \left [
\hat A(t), \; H(t) \right ] \right \rgl
\qquad \left ( \frac{\prt \hat A}{\prt t} = 0 \right ) \; .
\ee
The evolution equation is to be complemented by an initial condition
$A(0) = <\hat{A}(0)>$. It is the memory of such initial conditions
that can be retained in the representative ensemble characterizing
equilibrated finite systems.

\subsection{Equations in matrix representation}

Let $\{|n>\}$ be an orthonormal basis in the Hilbert space of
microstates $\mathcal{F}$. Define the matrix elements for the
statistical operator
\be
\label{21}
\rho_{mn}(t) \equiv \lgl m | \hat\rho(t) | n \rgl
\ee
and for the operator of an observable
\be
\label{22}
A_{mn}(t) \equiv \lgl m | \hat A(t) | n \rgl \;  .
\ee
The related matrix elements at zero time are
\be
\label{23}
\rho_{mn} \equiv \lgl m | \hat\rho | n \rgl = \rho_{mn}(0) \; ,
\qquad
A_{mn} \equiv \lgl m | \hat A | n \rgl = A_{mn}(0) \; .
\ee
Then the average (\ref{15}) can be represented as
\be
\label{24}
 \lgl \hat A(t) \rgl = \sum_{mn} \rho_{mn}(t) A_{nm} =
\sum_{mn} \rho_{mn} A_{nm}(t) \; .
\ee

This average, in general, contains nondiagonal terms. Of course,
if one chooses as the basis the set of the eigenvectors of $\hat{A}$,
then the sum would contain only the diagonal terms. However, other
operators, noncommuting with $\hat{A}$, will have nondiagonal terms.
When one discusses equilibration, one should keep in mind not just
one operator but the set of operators from the algebra of local
observables $\mathcal{A}$. Equilibration, by definition, presupposes
that it is the whole statistical state $<\mathcal{A}>$ that
equilibrates, but not merely some of its components.

\subsection{Pure quantum states}

One often considers equilibration of quantum systems prepared in
a pure state described by a normalized wave function $|\psi(0)>$.
With a wave function $|\psi(t)>$, the statistical operator of a
pure state is
\be
\label{25}
 \hat\rho(t) = |\psi(t)\rgl \lgl \psi(t) | \; .
\ee
It is an idempotent operator, with the property
$\hat{\rho}^2(t) = \hat{\rho}(t)$. The wave function satisfies the
Schr\"{o}dinger equation
\be
\label{26}
i\; \frac{d}{dt} \; |\psi(t) \rgl = H |\psi(t) \rgl \;  .
\ee
Employing an expansion
\be
\label{27}
 |\psi(t) \rgl = \sum_n c_n(t) | n \rgl \; ,
\qquad c_n(t) \equiv \lgl n | \psi(t) \rgl \; ,
\ee
one gets the equation for the expansion functions
\be
\label{28}
 i \; \frac{d}{dt} \; c_n(t) = \sum_m H_{nm} c_m(t) \; ,
\ee
where $H_{mn} \equiv <m|H|n>$, complimented by an initial condition
$c_0 = c(0)$ and the normalization condition
$$
 \sum_n | c_n(t) |^2 = 1 \; .
$$
The matrix element $H_{mn}$ can be explicitly dependent on time.

The statistical operator (\ref{25}) enjoys the expansion
\be
\label{29}
 \hat\rho(t) = \sum_{mn} \rho_{mn}(t) | m \rgl \lgl n | \; ,
\ee
in which
$$
\rho_{mn}(t) = c_m(t) c_n^*(t) \;  .
$$

Average (\ref{24}) takes the form
\be
\label{30}
 \lgl \hat A(t) \rgl = \sum_{mn} \rho_{mn}(t) A_{nm} =
\lgl \psi(t) | \hat A | \psi(t) \rgl  \; .
\ee
The pure-state statistical operator (\ref{25}) is a particular
case of the general statistical operator. Because of this, it is
sufficient to consider the general case.

\section{Isolated quantum systems}

\subsection{Absence of absolute equilibrium}

Let us assume that a finite quantum system can be completely
isolated from its surrounding. Hence the Hamiltonian $H$ does not 
explicitly depend on time. Then the evolution operator is
\be
\label{31}
 \hat U(t) = e^{-i Ht} \qquad \left ( \frac{\prt H}{\prt t} = 0
\right )  \; .
\ee
It is admissible to take the set of the Hamiltonian eigenvectors,
defined by the eigenproblem
\be
\label{32}
H | n \rgl = E_n | n \rgl \;   ,
\ee
as the basis in $\mathcal{F}$. Then the matrix element (\ref{21})
becomes
\be
\label{33}
\rho_{mn}(t) = \rho_{mn} e^{-i\om_{mn}t} \;  ,
\ee
with the transition frequency
\be
\label{34}
 \om_{mn} \equiv E_m - E_n \; .
\ee
The operator average (\ref{24}) yields
\be
\label{35}
 \lgl \hat A(t) \rgl = \sum_{mn}
\rho_{mn} A_{nm} e^{-i\om_{mn}t } \; .
\ee
And the equation of motion (\ref{20}) reduces to
\be
\label{36}
 i \; \frac{d}{dt} \; \lgl \hat A(t) \rgl =
\sum_{mn} \rho_{mn} A_{nm} \om_{mn} e^{-i\om_{mn}t} \; .
\ee

It is evident that Eq. (\ref{35}) is a quasi-periodic function of
time. Hence, it cannot tend to a time-independent stationary state.
Any given initial state will reproduce itself after the Poincar\'{e}
recurrence time [81-84]. The temporal limit (\ref{4}) does not exist.
{\it A finite quantum system, completely isolated from its surrounding,
does not have an absolute equilibrium in the sense of limit (\ref{4})}.

An exception could be the case, when the system would be initially 
prepared in a pure state, with a given Hamiltonian eigenvector $|j>$, 
when $\rho_{mn} = \delta_{mj} \delta_{nj}$. Then the average never 
changes with time:
\be
\label{37}
\lgl \hat A(t) \rgl = A_{jj} \qquad
(\rho_{mn} = \dlt_{mj} \dlt_{nj} ) \;  .
\ee
However, the dynamics of the system with this initial state is not
structurally stable. If the initial condition slightly deviates from
the pure state $|j>$ or there is a perturbation of the Hamiltonian,
so that the initial condition gives
$$
 \rho_{mn} = \dlt_{mj} \dlt_{nj} + \dlt \rho_{mn} \;  ,
$$
then the average
$$
\lgl \hat A(t) \rgl = A_{jj} + \sum_{mn} (\dlt \rho_{mn} )
A_{nm} e^{-i\om_{mn} t}
$$
again becomes quasi-periodic.

\subsection{Possibility of equilibrium on average}

Von Neumann [85] suggested that quantum systems could exhibit a kind
of ergodic behavior, similarly to that of classical systems, when the
time average of the expectation value of an observable coincides with
its ensemble-averaged value. For this purpose, one considers the time
average
\be
\label{38}
\overline{\lgl \hat A(t) \rgl } \equiv \lim_{\tau\ra\infty}\;
\frac{1}{\tau} \; \int_0^\tau \lgl \hat A(t) \rgl \; dt \;  .
\ee
For an isolated system, this gives
\be
\label{39}
\overline{\lgl \hat A(t) \rgl } = 
\sum_{mn} \rho_{mn} A_{nm} \Dlt(\om_{mn} ) \;   ,
\ee
where
$$
\Dlt(\om) \equiv  \lim_{\tau\ra\infty}\;
\frac{1}{\tau} \; \int_0^\tau e^{-i\om t} \; dt \; ,
$$
which gives
\begin{eqnarray}
\nonumber
\Dlt(\om) = \left \{ \begin{array}{ll}
1 \; , & ~ \om = 0 \\
0 \; , & ~ \om \neq 0 \; . 
\end{array} \right.
\end{eqnarray}
If the eigenenergies of the Hamiltonian are not degenerate, then
\be
\label{40}
\Dlt(\om_{mn} ) \equiv \lim_{\tau\ra\infty} \; \frac{1}{\tau}\;
\int_0^\tau e^{-i\om_{mn} t} \; dt = \dlt_{mn} \;  .
\ee
On usually assumes this {\it nondegeneracy condition}, which yields
\be
\label{41}
\overline{\lgl \hat A(t) \rgl } = 
\sum_n \rho_{nn} A_{nn} \;  .
\ee
Hamiltonians, enjoying nondegenerate spectra, are termed typical
or generic.

Von Neumann assumed that, under some conditions, the right-hand side
of Eq. (\ref{41}) would correspond to the average over an equilibrium
ensemble. In that sense, Eq. (\ref{41}) would play the role of an
ergodic relation, similar to that in classical systems.

It may happen, however, that, even when this ergodic-type relation
is valid, but system fluctuations around the averaged value are so
large that it is senseless to talk about equilibrium. In order that
it would be possible to state that a quantum systems spends the most
of its time close to the averaged value, having not so many and not
so large deviations from it, it is necessary to analyze the strength
of the system fluctuations. For an operator $\hat{A}$, these
fluctuations are quantified by the dispersion
\be
\label{42}
\sgm_A^2 \equiv \overline{\lgl \hat A(t) \rgl^2 } \; - \;
\left ( \; \overline{\lgl \hat A(t) \rgl }  \;  \right )^2  \;  ,
\ee
which is also called the variance.

Considering the dispersion, one usually assumes that, in addition
to the nondegeneracy condition (\ref{40}), the
{\it nonresonance condition}
\be
\label{43}
\lim_{\tau\ra\infty} \; \frac{1}{\tau}\;
\int_0^\tau e^{-i(\om_{mn}+\om_{kl}) t} \; dt =
\dlt_{mn} \dlt_{kl} + \dlt_{ml} \dlt_{nk} -
\dlt_{mn} \dlt_{nk} \dlt_{kl} 
\ee
holds true. If so, it immediately follows that
\be
\label{44}
 \sgm_A^2 = \sum_{m\neq n} | \rho_{mn} A_{nm} |^2 \; .
\ee
One also employs the inequality
$$
\sum_n \rho_{nn}^\al \; \leq \; 
( \max_n \rho_{nn} )^{\al-1} \;  ,
$$
being valid for any real $\alpha > 1$.

Another important point is the {\it Chebyshev inequality} stating
that for any random variable $x$, with average $\bar{x}$ and
variance $\sigma^2$, and any given $\varepsilon > 0$, the
probability that $x$ deviates from $\bar{x}$ by more than
$\varepsilon$ satisfies the inequality
\be
\label{45}
P\left ( | x - \overline x | \; > \; \ep \right ) \; < \;
\left ( \frac{\sgm}{\ep} \right )^2 \;   .
\ee
This inequality allows for the estimation of the operator variance
(\ref{44}) characterizing the deviations from the operator average
[86-88].

For the operator average $<\hat{A}>$, assuming that the experimental
resolution $\delta_A$ of measuring the observable, associated with
the operator $\hat{A}$, is sufficiently small, and that
$max_n \rho_{nn} \ll 1$, Reimann [87,88] obtains
\be
\label{46}
 P \left ( \left | \lgl \hat A(t) \rgl \; - \; 
\overline{ \lgl \hat A(t) \rgl } \right | \; \geq \; 
\dlt_A \right )  \; \leq \; 
\left ( \frac{\sgm_A}{\dlt_A} \right )^2 \; .
\ee
This inequality tells us that, under some conditions, the deviations
of the observable quantity $<\hat{A}>$ from the time-averaged value
$\overline{\lgl \hat{A}\rgl }$ could be relatively small. Hence the 
system would spend its major time near the time-averaged value.

The assumption that $max_n \rho_{nn} \ll 1$ presupposes a uniform
distribution over many degrees of freedom, which is equivalent to
ascribing to the system a high effective temperature. Such an
assumption loses its validity at low effective temperatures, when
it can be that $\rho_{nn} \sim 1$, for instance, when there is
Bose-Einstein condensation.

\subsection{Eigenstate thermalization on average}

If the initial statistical operator corresponds to a fixed pure
state $|j>$ that is a Hamiltonian eigenstate then, according to
Eq. (\ref{37}), the observable quantities do not change in time,
\be
\label{47}
 \lgl \hat A(t) \rgl = A_{jj} = \overline{\lgl \hat{A}\rgl } \; ,
\ee
which is evident, since the Hamiltonian eigenstate is a stationary
state. It is reasonable to expect that there could exist a set of
states around a given eiegenstate from which the system would
equilibrate on average to a state close to that eigenstate.

Let us fix the eigenstate $|j>$, whose energy is $E_j$. And let
us consider an energy shell
\be
\label{48}
\mathbb{E}_j \equiv \{ E_n : \; | E_n - E_j| < \Dlt E_j \}
\ee
of energies deviating from $E_j$ not more than by $\Delta E_j$.
The energy shall is characterized by the indicator function
\begin{eqnarray}
\label{49}
\xi_j(E_n) \equiv \left \{ \begin{array}{ll}
1 \; , & ~ E_n \in \mathbb{E}_j \\
0 \; , & ~ E_n \not\in \mathbb{E}_j \; .
\end{array} \right.
\end{eqnarray}
If we take for an initial condition an eigenvector with the energy
in the energy shell (\ref{48}), then we have
\be
\label{50}
\rho_{nn} \equiv p_{nj} \xi_j(E_n) \;   ,
\ee
with the normalization condition
$$
\sum_n \rho_{nn} = \sum_n p_{nj} \xi_j(E_n) = 1 \;  .
$$

The variation of the diagonal matrix elements of an operator, in
the range of the energy shell, is smaller than
\be
\label{51}
\Dlt A_j \equiv \max_{E_n\in\mathbb{E}_j} A_{nn} -
\min_{E_n\in\mathbb{E}_j} A_{nn} \; ,
\ee
which is a non-negative quantity. The values $A_{nn}$ lie in the
vicinity of the average
\be
\label{52}
 A_j^* \equiv 
\frac{\sum_n \xi_j(E_n)A_{nn}}{\sum_n\xi_j(E_n)} \;  .
\ee
It is always possible to choose so narrow energy shell that
\be
\label{53}
\left | \frac{\Dlt A_j}{A_j^*} \right | \ll 1 \;   .
\ee
Then the elements $A_{nn}$ are almost constant inside the chosen
energy shell, so that one can invoke the theorem of average
\be
\label{54}
 \sum_n p_{nj} \xi_j(E_n) A_{nn} \simeq 
A_j^* \sum_n p_{nj} \xi_j(E_n) = A_j^* \; .
\ee
Therefore, for any normalized weight $p_{nj}$, we have
\be
\label{55}
 \overline{\lgl \hat A(t) \rgl } \simeq A_j^* \; .
\ee

Equality (\ref{55}) implies that there exists an attraction
basin in the set of initial states, from which there happens an
approximate equilibration on average to the value $A_j^*$.
Since, under condition (\ref{53}), the weight $p_{nj}$ can be
taken as an arbitrary normalized distribution, it can be chosen
to be a simple uniform distribution
\be
\label{56}
p_{nj} = \frac{1}{\sum_n \xi_j(E_n)} \; .
\ee
This allows us to rewrite average (\ref{52}) in the form
\be
\label{57}
 A_j^* = \sum_n p_{nj} \xi_j(E_n) A_{nn} \; ,
\ee
which corresponds to the Gibbs microcanonical ensemble. Such an
equilibration can be called the eigenstate thermalization on average.
Different variants of this type of thermalization have been
considered in Refs. [89-92].

It is worth emphasizing that this thermalization is approximate, it
happens on average, initial states are assumed to be pure and
sufficiently close to a fixed eigenstate, the energy shell has to be
rather narrow, with the width of this shell depending on the
considered observable, and also the width can depend on the statistics
of the system components.

\subsection{Density of quantum states}

The properties of a quantum system are governed by its Hamiltonian
spectrum, whose form should also be important for quantum system
equilibration. Because of the existence of the correspondence principle,
one often compares quantum systems with classical ones. The equilibration
of the latter depends on whether the system is integrable or not.
Integrable classical systems do not equilibrate. For equilibration to
happen, the system has to be nonintegrable.

The notion of integrability is well defined for classical systems. If
such a system, with $n$ degrees of freedom, possesses $n$ independent
functions in involution (mutually Poisson commuting), then the system
can be integrated up to quadratures [52]. This is known as Liouville
theorem.

Quantum integrable systems in $n$ dimensions are often defined analogously
by requiring the existence of $n$ mutually commuting operators. However,
there is no a theorem equivalent to the classical Liouville theorem,
and there is nothing resembling the reduction to quadratures. In quantum
mechanics, an integrable system is understood as that for which the
spectral eigenproblem can be solved exactly [93]. The class of such
exactly solvable quantum problems is very limited.

Equilibration in closed classical systems is usually accompanied by
the appearance of chaos [52]. The notion of quantum chaos is not well
defined. One often resorts to the correspondence principle, telling
that quantum chaos can exist, provided that the corresponding classical
system is chaotic [94]. The latter requires that the system be
nonintegrable. However, classical chaos does not necessarily imply
quantum chaos [95]. Quantum chaos is assumed to be related to the
properties of the energy spectra.

Gutzwiller [96] proposed that the spectra of integrable and
nonintegrable quantum systems should be qualitatively different,
which would result in the qualitative difference of the density of
states
\be
\label{58}
\rho(E) = \sum_n \dlt(E-E_n) \;   .
\ee
It has been suggested [96-100] that, generally, the spectra of quantum
systems consist of two parts, regular and irregular, the regular part
being almost equidistant, while the irregular part being random.
Respectively, the density of states can be represented as a sum of
two terms, regular and irregular [94,96,101,102],
\be
\label{59}
\rho(E) = \rho_{reg}(E) + \rho_{irr}(E) \;  .
\ee
The regular term, in the semiclassical approximation, reads as
\be
\label{60}
\rho_{reg}(E) = \frac{1}{\Gm(d/2)} \; 
\left ( \frac{m}{2\pi}\right )^{d/2}  \;
\int \Theta(E-U(\br)) [ E - U(\br) ]^{(d-2)/2} \; d\br \; ,
\ee
where $d$ is the space dimensionality and $U({\bf r})$ is an external
potential. The semiclassical density of states (\ref{60}) is widely
used for describing cold trapped atoms (see., e.g., [103]). If the
external potential is a homogeneous function of degree $\alpha$,
such that
$$
 U(\lbd\br) = \lbd^\al U(\br) \;  ,
$$
then the regular term is
$$
\rho_{reg}(E) \cong \frac{(2+\al)d}{2\al} \;
E^{(2+\al)d/2\al-1} \;  .
$$
In particular, for a harmonic trap, when $\alpha = 2$, one has
$$
\rho_{reg}(E) \cong E^{d-1} d \qquad (\al=2) \;  .
$$
The irregular term of the density of states can be represented as
\be
\label{61}
 \rho_{irr}(E) \cong {\rm Im} \sum_n A_n \exp\left \{
i \left ( S_n \; - \; \frac{\pi}{2}\; \bt_n \right ) 
\right \} \; ,
\ee
where the sum is over semiclassical trajectories, $S_n$ is an
action integral, and $\beta_n$ is the Maslov index [94,96,104].
The amplitude $A_n$ and the action integral $S_n$ are continuous
functions of the energy levels. The irregular term of the
density of states is a fastly oscillating function of energy.

\subsection{Statistics of level spacing}

The most evident difference between integrable and nonintegrable
quantum systems is seen in the statistics of their level spacing.
One considers the reduced spectrum spacing
\be
\label{62}
s \equiv \frac{\Dlt E}{\lgl \Dlt E \rgl } \;   ,
\ee
where $\Delta E$ is the energy difference between the nearest
energy levels, arranged so that $\Delta E$ be positive, hence
$s \in [0,\infty)$.

For {\it integrable} quantum systems, the level spacing is
random [99,105], being described by the Poisson distribution
\be
\label{63}
P(s) = e^{-s} \;  .
\ee
The latter has the mean $\lgl s\rgl=1$ and dispersion $\sgm_s^2=1$. 
The most probable energy gap is $s_P = 0$.

{\it Nonintegrable} quantum systems are characterized by the
Wigner [106] distribution of level spacing
\be
\label{64}
 P(s) = \frac{\pi}{2} \; s \; \exp \left ( -\; 
\frac{\pi}{4} \; s^2 \right ) \;  .
\ee
This has the mean $\lgl s\rgl = 1$ and dispersion $\sgm_s^2=0.273$. 
The most probable energy gap is
\be
\label{65}
 s_W = \sqrt{ \frac{2}{\pi} } \; = \; 0.798 \; .
\ee
The existence of the nonzero gap (\ref{65}) is termed the level
repulsion.

Thus, integrable and nonintegrable quantum systems display essentially
different statistics of their level spacing. Therefore equilibration
can proceed differently for these systems. However, as is discussed
in the previous subsections, integrable quantum systems can exhibit
equilibration on average to a representative ensemble, though, may be,
slower than nonintegrable systems. This is contrary to classical
integrable systems that do not equilibrate at all.

\section{Characteristic time scales}

\subsection{Interaction time}

The process of equilibration of quantum systems from a strongly
nonequilibrium state goes through several stages characterized
by the corresponding time scales [7,62,82,107].

The shortest characteristic time is the time during which two
particles interact with each other. For atomic gases, this time
is of order $t_{int} \sim a_s/v$, with $a_s$ being a scattering
length and $v \sim 1/m a_s$, atomic velocity, where $m$ is mass.
Thence, the interaction time reads as
\be
\label{66}
 t_{int} = ma_s^2 \; .
\ee
For Hubbard-type models with the on-site interaction $U$, the
interaction time is
\be
\label{67}
 t_{int} = \frac{1}{U} \; .
\ee
Generally, the interaction time is inverse to the typical strength
of the strongest particle interactions. This time defines the first
stage of relaxation
\be
\label{68}
 0 < t < t_{int} \qquad (interaction \; stage) \;  ,
\ee
when particles move being yet not correlated with each other.

\subsection{Local-equilibrium time}

After the interaction time, particles start developing mutual
correlations that are growing till the local-equilibrium time
that, for atomic gases, is of the order $t_{loc} \sim \lambda/v$,
where $\lambda \sim 1/ \rho a_s^2$ is the mean free path and
$\rho$ is an average particle density. Then the local-equilibrium
time is
\be
\label{69}
t_{loc} = \frac{m}{\rho a_s} \;   .
\ee
The interval of time between the interaction time and
local-equilibrium time is the kinetic stage:
\be
\label{70}
t_{int} < t < t_{loc} \qquad (kinetic \; stage) \; ,
\ee
when the motion can be described by kinetic equations.

\subsection{Heterophase time}

After the local-equilibrium time $t_{loc}$, the evolution enters
hydrodynamic stage that can be subdivided into substages, when
the system dynamics crosses a phase transition line. If the system
starts from one phase, but is quenched into the conditions
supporting another phase, then, after $t_{loc}$, there appear the
nuclei of the new phase, which are also called clusters, or droplets,
of this new phase, or mesoscopic topological defects. These are
randomly intermixed in space with the regions of the former phase.
Frenkel [108] called such systems heterophase. The nuclei of the
competing phase can be of various shapes. In particular, they can
be in the form of vortices [26,27,56]. Such a heterophase mixture
exists during the heterophase stage
\be
\label{71}
t_{loc} < t < t_{het} \qquad ( heterophase \; stage) \;   ,
\ee
till the heterophase time $t_{het}$, when the system becomes uniform,
with the more stable phase filling the system volume. Equilibration
passing through the heterophase stage has been observed, e.g., in
experiments [20,22,24]. The heterophase stage can be of different 
duration, with $t_{het}$ being defined by the system properties [56,62]. 
In some cases, the heterophase time can be so long that the system can 
be treated as quasi-equilibrium during the whole stage (\ref{71}), 
which happens for many condensed-mater systems [62,107,109-118].

\subsection{Relaxation time}

The second substage of the hydrodynamic stage is that during which the
system relaxes to an equilibrium or quasi-equilibrium state. This is the
relaxation stage that lasts in the interval of time
\be
\label{72}
 t_{het} < t < t_{rel} \qquad (relaxation \; stage) \;  ,
\ee
if there exists the heterophase stage (\ref{71}). If the latter does not
exist, then the relaxation stage starts from the local-equilibrium time
(\ref{69}). The heterophase stage may be absent, when the initial
nonequilibrium conditions correspond to the same phase to which the
system is assumed to relax, so that no transition line is crossed [119].

In numerical simulations for the Hubbard model, corresponding to atoms
in optical lattices, the relaxation time was found to be
\be
\label{73}
t_{rel} = \frac{1}{J} \;   ,
\ee
for both bosons [46,47] as well as fermions [48]. Introducing [18] the
effective mass
\be
\label{74}
 m^* \equiv \frac{1}{2Ja^2} \;  ,
\ee
where $a$ is the distance between the nearest-neighbor sites, transforms
the relaxation time (\ref{73}) to the form
\be
\label{75}
t_{rel} = 2m^* a^2 \;  .
\ee
The effective mass, of course, can be very different from the
atomic mass $m$.

For quasi-one-dimensional Bose gases with local interactions, the
relaxation rates were calculated by Mazets and Schmiedmayer [120].
The rate of populating the radially excited modes by pairwise atomic
collisions, at $T < \omega_{\perp}$ was found to be
\be
\label{76}
 \Gm_2 = \al \om_\perp \exp \left ( -\; \frac{2\om_\perp}{T}
\right ) \;  ,
\ee
with the parameter
\be
\label{77}
 \al \equiv \frac{Na_s^2}{2l_zl_\perp} \;  .
\ee
Here $N$ is the number of atoms in the trap,
$l_z\equiv 1/\sqrt{m\om_z}$ and $l_\perp \equiv 1/\sqrt{m\om_\perp}$ 
are the effective trap lengths in the longitudinal and transverse 
directions, while $\om_z$ and $\om_\perp$ are the related trap 
frequencies. The relaxation rate, due to three-body collisions [120], 
is
\be
\label{78}
 \Gm_3 = 2 \al^2 \om_\perp \;  .
\ee
The relaxation time is defined by the minimal inverse rate:
\be
\label{79}
 t_{rel} = \min \left \{ \frac{1}{\Gm_2} \; , \;
\frac{1}{\Gm_3} \right \} \;  ,
\ee
that is, by the maximal relaxation rate. Atomic correlations, described
by the two-body and three-body correlation functions $g_n(0)$ at zero
distance, can diminish [120] the relaxation rates to
\be
\label{80}
 \widetilde\Gm_n = \Gm_n g_n(0) \qquad (n=2,3) \;  ,
\ee
resulting in the increase of the relaxation time.

After the relaxation time, the system enters a quasi-equilibrium regime,
lasting till the recurrence time.

\subsection{Recurrence time}

As is seen from Eq. (\ref{35}), the operator average is quasi-periodic,
hence it always returns arbitrarily close to any given initial value 
after the recurrence time that can be estimated as
\be
\label{81}
 t_{rec} \approx \frac{2\pi}{\Dlt E} \;  ,
\ee
where $\Dlt E$ is the mean spacing between energy eigenvalues near 
$E$. The mean spacing is of the order $\Dlt E \sim E/\dlt_E$, with
$\dlt_E$ being the number of states in the interval $\Dlt E$. The
entropy, as is known, is related to $\delta_E$ as $S \sim \ln \delta_E$,
thence $\delta_E \sim \exp (S)$. For systems with $N \gg 1$, one has
$S \sim N$ and $E \sim \varepsilon N$, with $\varepsilon$ being energy per
atom. Therefore the recurrence time can be represented by the 
expression
\be
\label{82}
 t_{rec} = \frac{2\pi}{\ep N} \; e^N \;  .
\ee
If the reduced energy $\varepsilon$ defines the interaction time
$t_{int} \sim 1/ \varepsilon$, then
\be
\label{83}
t_{rec} = \frac{2\pi}{N} \; e^N t_{int} \;   .
\ee
The recurrence time, for large $N$, can be extremely long. The
quasistationary stage is associated with the time interval
\be
\label{84}
t_{rel} < t < t_{rec} \qquad (quasistationary \; stage) \;  .
\ee

Generally, the typical relations between the characteristic times
satisfy the inequalities
\be
\label{85}
 t_{int} < t_{loc} < t_{het} < t_{rel} < t_{rec} \; .
\ee
Numerical values of these time scales, for condensed-matter systems
and atomic gases, are given in Refs. [7,62].

\section{Nonisolated quantum systems}

\subsection{Quantum system in a bath}

If a finite quantum system is not isolated, one needs to treat the
coupled pair of the system and its surrounding, called bath. The
composite system is characterized by the space of microstates
\be
\label{86}
\cH_{AB} \equiv \cH_A \bigotimes \cH_B \;   ,
\ee
being the tensor product of the Hilbert space $\cH_A$, related to 
the studied system, and of the Hilbert space $\cH_B$, related to 
the bath. The dimension of the composite space is
\be
\label{87}
 {\rm dim} \cH_{AB} = d_A d_B \;  ,
\ee
where the factor space dimensions are
\be
\label{88}
d_A \equiv {\rm dim} \cH_A \; , \qquad
d_B \equiv {\rm dim} \cH_B \;   .
\ee

The total Hamiltonian is the sum
\be
\label{89}
 H_{AB} = H_A \bigotimes \hat 1_B + \hat 1_A \bigotimes H_B 
+ H_{int} \;  ,
\ee
where $\hat{1}_A$ and $\hat{1}_B$ are the unity operators defined on
$\mathcal{H}_A$ and $\mathcal{H}_B$, respectively. To simplify the
notation, one usually omits the unity operators, writing
\be
\label{90}
 H_{AB} = H_A + H_B + H_{int} \;  ,
\ee
which implies Eq. (\ref{89}).

The statistical operator of the composite system is denoted by
$\hat{\rho}_{AB}(t)$, which can be pure or not. One introduces
partial statistical operators
\be
\label{91}
 \hat\rho_A(t) \equiv {\rm Tr}_B  \hat\rho_{AB}(t) \; , \qquad
\hat\rho_B(t) \equiv {\rm Tr}_A  \hat\rho_{AB}(t) \; ,
\ee
in which $Tr_A$ means the trace over the space $\mathcal{H}_A$
and $Tr_B$, the trace over $\mathcal{H}_B$.

Observable quantities are related to the system defined on
$\mathcal{H}_A$. The expectation value of an operator $\hat A$
of a local observable, acting on $\mathcal{H}_A$, is
\be
\label{92}
 \lgl \hat A(t) \rgl \equiv {\rm Tr}_{AB}  \hat\rho_{AB}(t) 
\hat A = {\rm Tr}_A \hat\rho_A(t) \hat A \;  ,
\ee
where $Tr_{AB}$ is the trace over the space $\mathcal{H}_{AB}$.

\subsection{Distance between statistical operators}

If one is interested in the equilibration on average, one needs
to consider the time-averaged statistical operator of the composite
object,
\be
\label{93}
 \overline\rho_{AB} \equiv \lim_{\tau\ra\infty} \; 
\frac{1}{\tau}\;  \int_0^\tau \hat\rho_{AB}(t) \; dt \; ,
\ee
and the time-averaged partial statistical operators
$$
\overline\rho_{A} \equiv \lim_{\tau\ra\infty} \; 
\frac{1}{\tau}\;  \int_0^\tau \hat\rho_{A}(t) \; dt \; ,
$$
\be
\label{94}
 \overline\rho_{B} \equiv \lim_{\tau\ra\infty} \; 
\frac{1}{\tau}\;  \int_0^\tau \hat\rho_{B}(t) \; dt \;  .
\ee

The difference between two statistical states, described by the
operators $\hat{\rho}_1$ and $\hat{\rho}_2$, defined on the same
Hilbert space $\mathcal{H}$, can be characterized by the
Hilbert-Schmidt trace distance
\be
\label{95}
 {\rm dist} \left [ \hat\rho_1,\; \hat\rho_2 \right ] 
\equiv  {\rm Tr}_\cH 
\sqrt{\left ( \hat\rho_1 - \hat\rho_2\right )^2} \;  .
\ee
Respectively, the time-averaged distance is
\be
\label{96}
 \overline{\rm dist} \left [ \hat\rho_1, \; \hat\rho_2 \right ] 
\equiv  \lim_{\tau\ra\infty} \; \frac{1}{\tau}\;  \int_0^\tau 
{\rm dist} \left [ \hat\rho_1(t) ,\; \hat\rho_2(t) \right ] \; 
dt \; .
\ee

It has been shown [121-123] that
\be
\label{97}
\overline{\rm dist} \left [\hat\rho_A(t), \; \overline{\hat\rho}_A 
\right ] \leq \sqrt{d_A\left ( {\rm Tr} \overline\rho_B^2\right ) } \;  .
\ee
In the proof, one uses the Jensen inequality, telling that for
a concave function, for which $d^2f(x)/dx^2 \leq 0$, one has
$\bar{f}(x) \leq f(\bar{x})$, and the inequality
$$
 {\rm Tr}_A \sqrt{\left ( \hat\rho_1 - \hat\rho_2\right )^2 }
\leq 
\sqrt{d_A{\rm Tr} \left ( \hat\rho_1 - \hat\rho_2 \right )^2} \; .
$$

If the bath is large, with a great number of uniformly distributed
states, such that
\be
\label{98}
 {\rm Tr} \overline\rho_B^2 \ll \frac{1}{d_A}\; ,
\ee
then the system equilibrates on average in the sense of the small
trace distance (\ref{97}).

\subsection{Decoherence induced by environment}

For the problem of equilibration, a large bath, representing an
environment in thermodynamic limit, is of special interest. The
standard definition of thermodynamic limit for a system of $N$
particles in a volume $V$ reads as
$$
 N \ra \infty \; , \qquad V \ra \infty \; , \qquad
\frac{N}{V} \ra const \;  .
$$
This definition, however, is not applicable for nonuniform quantum
systems confined in an external potential, such as trapped atoms,
since the system volume $V$ is not well defined in this case. A
general definition of thermodynamic limit, which is valid for
arbitrary systems, including confined ones, is given as follows [18,103].
For any extensive observable quantity $A_N \equiv <\hat{A}>$,
corresponding to a system of $N$ particles, it should be:
$$
 N \ra \infty \; , \qquad A_N \ra \infty \; , \qquad
\frac{A_N}{N} \ra const \;   .
$$
As is obvious, for a uniform system, where the volume $V$ is well
defined, the latter definition reduces to the standard one.

But we are interested in thermodynamic limit for a bath of volume
$V_B$ and described by an effective number of degrees of freedom $N_B$,
while the considered quantum system is kept finite, with a finite number
of atoms $N$. Then thermodynamic limit for the bath can be defined [124]
in the standard way
\be
\label{99}
N_B \ra \infty \; , \qquad V_B \ra \infty \; , \qquad
\frac{N_B}{V_B} \ra const \;   .
\ee
The role of the bath can be played by other parts of the same physical
system, where a finite part of it is fixed. In that case, thermodynamic
limit is applied to the whole system [125].

The effect of decoherence assumes that at long times the average
(\ref{92}) yields
\be
\label{100}
 \lim_{t\ra\infty} \; \lim_{N_B\ra\infty} \;
\lgl \hat A(t) \rgl = \sum_n \rho_{nn}^* A_{nn} \; ,
\ee
where the bath thermodynamic limit (\ref{99}) is implied. The
thermodynamic and time limits do not commute with each other.
Decoherence, as is defined in Eq. (\ref{100}), also means
equilibration, according to definition (\ref{4}).

Decoherence, as a general phenomenon, is supposed to hold for any
operator $\hat{A}$ from the algebra of local observables. Condition
(\ref{100}) may be trivially valid for some operators that commute
with the system Hamiltonian $H_A$, which has nothing to do with
decoherence.

The reduction of an operator average to the diagonal sum (\ref{100})
is called decoherence or dephasing [126,127] because of the evident
reason. In quantum mechanics, the nondiagonal terms characterize
interference effects that are typical of coherent processes. In
radiating systems, such nondiagonal terms are responsible for the
appearance of coherent radiation [71-80]. Hence, the disappearance
of the coherent terms is nothing but decoherence.

The properties of the environment are often characterized by random
distributions. For examples, expanding the wave function of the
composite system (system plus bath) over a basis, with the real and
imaginary parts of the expansion coefficients being independent
Gaussian random variables with zero mean, one comes to the reduced
statistical operator for the system, having canonical form [128,129].

Assuming a random distribution of the wave-function expansion
coefficients is equivalent to assuming infinite number of degrees of
freedom of environment. It is therefore possible, to impose the
condition of randomness on the wave function of the considered system,
without explicitly invoking a bath [130-132].

Similarly, equilibration and decoherence occur, if one assumes that
the system correlation functions exponentially fastly decay [133,134].
This assumption is analogous to the random-phase averaging and can be
interpreted as a random influence of environment.

\section{Quasi-isolated quantum systems}

\subsection{Absence of complete isolation}

The assumption of the possibility of an absolute system isolation is,
evidently, an idealization of the real situation. In reality, no one
system can be completely isolated. There always exists some random
influence of environment, though may be weak. Moreover, the notion of
absolute isolation is self-contradictory, since in order to state that
the given system has been isolated during a period of time, it is
necessary to accomplish a series of measurements with this system for
proving that the system has really been isolated during that time
span [60,135,136]. But each measurement, even if it is a nondemolition
measurement [137], disturbs the system, making it not isolated.
Measuring devices play the role of environment acting on the system.
And, vice versa, the influence of environment is analogous to the
action of measuring procedures.

In computer modeling, the role of external noise is played by numerical
errors accompanying calculations. With time, there is the accumulation
of errors, analogous to the increasing influence on the system of random
environment. Thus, a complete isolation is merely a theoretical
abstraction not existing in nature. There can exist only quasi-isolated
systems.

\subsection{Properties of quasi-isolated systems}

To formulate rigorously the notion of quasi-isolated systems, let us
proceed to mathematical definitions. Let $H_A$ be a system Hamiltonian
defined on the Hilbert space $\mathcal{H}_A$ and let $H_B$, defined on
$\mathcal{H}_B$, be the Hamiltonian of surrounding. The total space of
microstates for the composite object "system plus environment" is the
tensor product
\be
\label{101}
 \cH_{AB} \equiv \cH_A \bigotimes \cH_B \;  .
\ee
The Hamiltonian of the composite object is
\be
\label{102}
H_{AB} = H_A + H_B + H_{int}\;  ,
\ee
where $H_{int}$ characterizes interactions between the system and
environment and where the unity operators, for the brevity of notation,
are omitted.

It is reasonable to call a system quasi-isolated, when the matrix
representation for the operators of observable quantities, in the
basis of the eigenvectors of the system Hamiltonian $H_A$, is not
changed under the influence of the environment, so that this matrix
representation coincides with the matrix representation of these
operators in the basis of the eigenvectors of the total Hamiltonian
(\ref{102}). This means that $H_A$ is to be the integral of motion
with respect to $H_{AB}$,
\be
\label{103}
 [ H_A , \; H_{AB} ] = 0 \;  .
\ee
Condition (\ref{103}) is the mathematical definition of a
{\it quasi-isolated system}. The latter also implies the commutation
relation
\be
\label{104}
[ H_A , \; H_{int} ] = 0 \;  .
\ee
The property of quasi-isolatedness defines nondestructive, or
nondestroying, action of environment on the system, similarly to
the minimally disturbing action of measurements [60, 135,136].
This property should not be confused with the definition of
nondemolition measurements for a given observable [137], where
the commutation of the operator of the measured quantity with the
total Hamiltonian is required: $[\hat{A},H_{AB}] = 0$.

When the system Hamiltonian satisfies the eigenproblem
\be
\label{105}
H_A | n \rgl = E_n | n \rgl \;  ,
\ee
and the system is quasi-isolated, then the eigenproblem for $H_{AB}$
has the form
\be
\label{106}
 H_{AB} | nk \rgl = ( E_n + \ep_{nk} ) | nk \rgl \;  ,
\ee
where the vector $|nk> \equiv |n> \otimes |k>$ pertains to the space
$\mathcal{H}_{AB}$.

For the statistical ensemble $\{\mathcal{H}_{AB}, \hat{\rho}_{AB} (t)\}$,
the evolution of the statistical operator is given by the law
\be
\label{107}
 \hat\rho_{AB}(t) = \hat U_{AB}(t) 
\hat\rho_{AB}(0) \hat U_{AB}^+(t) \;  .
\ee
If $H_{AB}$ does not explicitly depend on time, then
\be
\label{108}
 \hat U_{AB}(t) = \exp( -i H_{AB} t ) \;  .
\ee
As an initial condition for the evolution of the statistical operator,
we accept
\be
\label{109}
 \hat\rho_{AB}(0) =\hat\rho_A(0) \bigotimes \hat\rho_B(0) \;  ,
\ee
which means that the interaction between the considered system and
the bath is switched on at $t = + 0$. This initial condition is not 
principal, but it is assumed here merely for simplicity.

\subsection{Equilibration of quasi-isolated systems}

We are interested in the temporal behavior of operators of local
observables, associated with the system. Say, we consider the average
of an operator $\hat{A}$ defined on $\mathcal{H}_A$. For this average,
we have the expression
\be
\label{110}
 \lgl \hat A(t) \rgl \equiv {\rm Tr}_{AB} \hat\rho_{AB}(t) \hat A
= {\rm Tr}_A \hat\rho_A(t) \hat A \; ,
\ee
in which
\be
\label{111}
\hat\rho_A(t) \equiv {\rm Tr}_B \hat\rho_{AB}(t) \; .
\ee
In the matrix representation, this yields
\be
\label{112}
 \lgl \hat A(t) \rgl = \sum_{mn} \rho_{mn}^A(t) A_{nm} \;  ,
\ee
where
\be
\label{113}
 \rho_{mn}^A(t) = \lgl m | \hat\rho_A(t) | n \rgl =
\sum_k \lgl mk | \hat\rho_{AB}(t) | nk \rgl \;  .
\ee
In view of the evolution operator (\ref{108}), Eq. (\ref{113})
can be written as
\be
\label{114}
 \rho_{mn}^A(t) = \sum_k \rho_{mnk}^{AB} \exp \{
- i( \om_{mn} + \ep_{mnk} ) t \} \;  ,
\ee
with the notation
\be
\label{115}
\rho_{mnk}^{AB} \equiv \lgl mk | \hat\rho_{AB}(0) | nk \rgl
\ee
and
\be
\label{116}
 \om_{mn} \equiv E_m - E_n \; , \qquad
\ep_{mnk} \equiv \ep_{mk} - \ep_{nk} \;  .
\ee

Thus for the average (\ref{110}), we get
\be
\label{117}
 \lgl \hat A(t) \rgl = \sum_{mn} \; \sum_k \rho_{mnk}^{AB}
A_{nm} \exp \{ -i (\om_{mn} + \ep_{mnk} ) t \} \;  .
\ee
Taking into account the initial condition (\ref{109}) yields
\be
\label{118}
 \rho_{mnk}^{AB} = \lgl m | \hat\rho_A(0) | n \rgl \; 
\lgl k | \hat\rho_B(0) | k \rgl \;  .
\ee
The latter, employing the notation
\be
\label{119}
 \rho_{mn} \equiv \lgl m | \hat\rho_A(0) | n \rgl \; , \qquad
p_k \equiv \lgl k | \hat\rho_B(0) | k \rgl \;  ,
\ee
reduces to
\be
\label{120}
 \rho_{mnk}^{AB} = \rho_{mn} p_k \;  .
\ee
The forms $\rho_{mn}$ and $p_k$ satisfy the normalization
conditions
\be
\label{121}
 \sum_n \rho_{nn} = 1 \; , \qquad \sum_k p_k = 1 \; .
\ee

Introducing the {\it decoherence factor}
\be
\label{122}
 D_{mn}(t) \equiv \sum_k p_k e^{-i\ep_{mnk} t} \;  ,
\ee
we come to the expression
\be
\label{123}
 \lgl \hat A(t) \rgl = \sum_{mn} \rho_{mn}(t) A_{nm}
D_{mn}(t) \;  ,
\ee
in which
\be
\label{124}
 \rho_{mn}(t) \equiv \rho_{mn} e^{-i\om_{mn}t} \;  .
\ee

Separating in summation (\ref{123}) the diagonal and nondiagonal
terms gives the average
\be
\label{125}
\lgl \hat A(t) \rgl = \sum_n \rho_{nn} A_{nn} +
\sum_{m\neq n} \rho_{mn}(t) A_{nm} D_{mn}(t) \;   ,
\ee
where we take into account that, according to definition (\ref{116}),
\be
\label{126}
 \ep_{nnk} = 0 \; .
\ee

Introducing the {\it density of states}
\be
\label{127}
 p_{mn}(\ep) \equiv \sum_k p_k \dlt(\ep -\ep_{mnk} ) \;  ,
\ee
satisfying the normalization condition
\be
\label{128}
 \int_{-\infty}^{\infty} p_{mn}(\ep) \; d\ep = 1\;  ,
\ee
transforms the decoherence factor (\ref{122}) to
\be
\label{129}
 D_{mn}(t) = \int_{-\infty}^{\infty} p_{mn}(\ep) 
e^{-i\ep t} \; d\ep \;  .
\ee
For the diagonal elements, we have 
$$
p_{nn}(\ep) = \sum_k p_k \dlt(\ep) \; , \qquad
D_{nn}(t) = 1 \;  .
$$
Therefore the decoherence factor is nontrivial only for the nondiagonal
elements of Eq. (\ref{129}).

Assume that the surrounding environment is large, such that the
spectrum of $H_{AB}$, with respect to the index $k$, can be treated
as continuous and the summation over $k$ can be replaced by integration.
Then the density of states (\ref{127}) may be considered as a smooth
function of $\varepsilon$. In that case, we can invoke the Riemann-Lebesgue
lemma [138] telling that, if $p_{mn}(\varepsilon)$ is a measurable
$L^1$-integrable function, then the time limit of Eq. (\ref{129}),
for $m \neq n$, is zero:
\be
\label{130}
\lim_{t\ra\infty} D_{mn}(t) = 0 \qquad (m\neq n) \;  .
\ee
Hence for average (\ref{125}), we obtain
\be
\label{131}
 \lim_{t\ra\infty} \lgl \hat A(t) \rgl = 
\sum_n \rho_{nn} A_{nn} \;  .
\ee

Therefore, a {\it quasi-isolated system in a large environment
equilibrates}. This has been proven here for an arbitrary
quasi-isolated system, which makes the given proof principally
different from the considerations, where the property of
quasi-isolatedness was not used, but instead, concrete systems
were studied. For instance, equilibration for quadratic Hamiltonians
was demonstrated [124,134]. Also, we have considered equilibration
in the strict sense of definition (\ref{4}), but not equilibration
on average [86-92], as described in Secs. 4.2 and 4.3.

\subsection{Examples of decoherence factors}

To illustrate explicitly admissible forms of the decoherence factor
(\ref{129}), let us consider several concrete cases.

When the density of states enjoys the Gaussian form
\be
\label{132}
p_{mn}(\ep) = \frac{1}{\sqrt{2\pi} \gm_{mn}}\;
\exp \left ( - \; \frac{\ep^2}{2\gm_{mn}^2} \right ) \; ,
\ee
then the decoherence factor is also Gaussian,
\be
\label{133}
 D_{mn}(t) = \exp \left ( -\; 
\frac{\gm_{mn}^2}{2} \; t^2 \right ) \;  .
\ee

For a uniform density of states
\be
\label{134}
 p_{mn}(\ep) = \frac{1}{2\gm_{mn} }\; \Theta(\gm_{mn}-\ep)
\Theta(\gm_{mn} + \ep) \;  ,
\ee
where $\Theta(\varepsilon)$ is a unit-step function, the decoherence
factor is
\be
\label{135}
D_{mn}(t) = \frac{\sin(\gm_{mn}t)}{\gm_{mn} t} \; .   
\ee

Assuming the Poisson distribution
\be
\label{136}
 p_{mn}(\ep) = \frac{1}{2\gm_{mn} } \;
\exp \left ( - \; \frac{|\ep|}{\gm_{mn} }\right ) \;  ,
\ee
yields the Lorentz-type decoherence factor
\be
\label{137}
 D_{mn}(t) = \frac{1}{1+(\gm_{mn}t)^2} \;  .
\ee

Conversely, for the Lorentz density of states
\be
\label{138}
 p_{mn}(\ep) = \frac{\gm_{mn} }{\pi(\ep^2+\gm_{mn}^2) } \;  ,
\ee
the decoherence factor is exponential,
\be
\label{139}
D_{mn}(t) = e^{-\gm_{mn}t} \;   .
\ee

In the above examples, the density of states is represented by
monotonic nonincreasing functions of $\ep$. For nonmonotonic
functions, the situation is similar, just formulas can become a
bit more cumbersome. For instance, if we take the density of
states in the Wigner form
\be
\label{140}
p_{mn}(\ep) = \frac{\pi|\ep|}{4\gm_{mn}^2} \;
\exp \left ( -\; \frac{\pi\ep^2}{4\gm_{mn}^2} \right ) \;  ,
\ee
then the decoherence factor reads as
\be
\label{141}
 D_{mn}(t) = 2 \int_0^\infty x e^{-x^2} \cos \left (
\frac{2\gm_{mn}t )}{\sqrt{\pi} } \; x \right ) \; dx \; .
\ee
The integral
$$
2 \int_0^\infty x e^{-x^2} \cos(qx) \; dx = 1 - q F \left (
\frac{q}{2} \right )
$$
is expressed through the Dawson function
$$
 F(z) = e^{-z^2} \int_0^z e^{x^2} \; dx \; .
$$
And for the decoherence factor we find
\be
\label{142}
D_{mn}(t) = 1 \; - \; 
\frac{2\gm_{mn}}{\sqrt{\pi} } \; t F \left (
\frac{\gm_{mn} t}{\sqrt{\pi} } \right ) \;  .
\ee
At large variable, the Dawson function behaves as
$$
F(z) \simeq \frac{1}{2z} + \frac{1}{4z^2} \qquad 
( z\ra \infty ) \;  .
$$
Consequently, the decoherence factor decays as
\be
\label{143}
 D_{mn}(t) \simeq -\; \frac{\pi}{2 \gm_{mn}^2t^2} \qquad
(t\ra \infty ) \; .
\ee

It is seen from these examples that the typical decoherence
time is
\be
\label{144}
 t_{dec} = \frac{1}{\gm} \;  ,
\ee
where $\gamma$ is an average $\gamma_{mn}$.

\subsection{Occurrence of rare events}

The property of asymptotic decoherence (\ref{130}) and, consequently,
of the equilibration limit (\ref{131}), are based on the use of
the Riemann-Lebesgue lemma assuming that the density of states
$p_{mn}(\varepsilon)$ is a measurable $L^1$ integrable function. By its
definition, the density of states is a non-negative function that is
normalized as in Eq. (\ref{128}). Hence, it is always $L^1$ integrable.
But it may happen that it can contain nonmeasurable parts. Then the
properties of decoherence and equilibration are not guaranteed.

For instance, the density of states can be formed of two parts,
\be
\label{145}
 p_{mn}(\ep) = \lbd_1 p_{mn}^{rel}(\ep) +
\lbd_2 p_{mn}^{osc}(\ep) \;  ,
\ee
in which the first term is a measurable function leading to a
relaxation process accompanied by decoherence and equilibration,
while the second term is a nonmeasurable function resulting in
decoherence oscillations, and where
\be
\label{146}
 \lbd_1 + \lbd_2 = 1 \qquad 
( \lbd_1 > 0 \; , \; \lbd_2 > 0 ) \;  .
\ee
Thence the decoherence factor is also a sum
\be
\label{147}
D_{mn}(t) = \lbd_1 D_{mn}^{rel}(t) + \lbd_2 D_{mn}^{osc}(t)
\ee
of the related terms, the first of which satisfies property 
(\ref{130}).

Let the second term of the density of states have the form
\be
\label{148}
p_{mn}^{osc}(\ep) = \frac{1}{2} \sum_j [ \dlt(\ep-\al_j) +
\dlt(\ep+\al_j) ] \;  ,
\ee
where $\alpha_j$ are real quantities. Then the oscillating part of the
decoherence factor is
\be
\label{149}
  D_{mn}^{osc}(t) = \sum_j \cos(\al_j t) \; .
\ee
Therefore the level of decoherence will oscillate, being a
quasi-periodic function characterized by the periods
\be
\label{150}
 t_j = \frac{2\pi}{\al_j} \;  .
\ee
Respectively, the operator average (\ref{123}) does not decohere
and the system does not equilibrate. However it may look as equilibrium
on average, keeping in mind temporal averaging.

If the parameters $\alpha_j$ are small, nonequilibrium oscillations
will be rare. But if they are strong, they cannot be neglected, since
their presence can essentially influence the system properties. Such
rare, but strong, events can support the existence of heterophase
fluctuations [62,107,109-118].

\section{Equilibration under nondestructive measurements}

\subsection{Definition of nondestructive measurements}

As has been mentioned above, measurements accomplished with the
considered system act similarly to the action of environment. The
main difference is that the measurement procedures, generally, 
depend on time, which makes the consideration more complicated.

Let the system be characterized by a Hamiltonian $H_A$, defined on
a Hilbert space $\mathcal{H}_A$, and a measuring device be described
by a Hamiltonian $H_D$, defined on a Hilbert space $\mathcal{H}_D$.
These Hamiltonians $H_A$ and $H_D$ are assumed to be time independent.

The total space of microstates is
\be
\label{151}
  \cH = \cH_A \bigotimes \cH_D \; .
\ee
The total Hamiltonian can be written as
\be
\label{152}
 H = H_A + H_M \;  ,
\ee
where the measurement Hamiltonian
\be
\label{153}
H_M = H_D + H_{int}
\ee
is the sum of the Hamiltonian $H_D$ of the measuring device and of
$H_{int}$ characterizing the interactions between the studied system
and the measuring device. The system Hamiltonian $H_A$ and the device 
Hamiltonian $H_D$ do not explicitly depend on time, but the 
interaction Hamiltonian $H_{int}$ is a function of time. Consequently, 
the measurement Hamiltonian is an explicit function of time, 
$H_M = H_M(t)$. As usual, we omit in the formulas the unity operators, 
keeping in mind that the rigorous expression for Hamiltonian (\ref{152}) 
would be
\be
\label{154}
H = H_A \bigotimes \hat 1_D + \hat 1_A \bigotimes H_D + 
H_{int} \;   .
\ee

We consider measurements that minimally disturb the system [60,135,136],
because of which they are named nondestructive. The observed system has 
to be quasi-isolated from the measuring device. A measurement is called
{\it nondestructive}, if the system Hamiltonian, as well as the device 
Hamiltonian do not disturb the properties of each other, being the 
integrals of motion:
\be
\label{155}
[ H_A, \; H] = 0 \qquad , [ H_D, \; H] = 0  \;  .
\ee
It follows from Eq. (\ref{155}) that
\be
\label{156}
 [ H_A, \; H_M ] = 0 \qquad  , [ H_D, \; H_M ] = 0 \;  . 
\ee

If the eigenproblem for $H_A$ has the form
\be
\label{157}
H_A | n \rgl = E_n | n \rgl \;   ,
\ee
then the measurement Hamiltonian satisfies the eigenproblem
\be
\label{158}
 H_M(t) | nk \rgl = \ep_{nk}(t) | nk \rgl \;  ,
\ee
where $|nk> \equiv |n> \otimes |k>$, with $\{|n>\}$ being a basis in
$\mathcal{H}_A$ and $\{|k>\}$, a time-independent basis in
$\mathcal{H}_D$.

The statistical operator for the composite system is
\be
\label{159}
 \hat\rho(t) = \hat U(t) \hat\rho(0) \hat U^+(t) \;  ,
\ee
in which the initial condition
\be
\label{160}
\hat\rho(0) = \hat\rho_A(0) \bigotimes \hat\rho_D(0)
\ee
is assumed. This means that measurements start at the moment
of time $t = + 0$.

In view of eigenproblem (\ref{158}), it follows that the measurement
Hamiltonian satisfies the Lyappo-Danilevsky condition
\be
\label{161}
\left [ H_M(t), \; \int_0^t H_M(t') \; dt' \right ]
= 0 \;   .
\ee
Since the system Hamiltonian $H_A$ does not explicitly depend on
time, the previous property (\ref{161}) gives
\be
\label{162}
 \left [ H(t), \; \int_0^t H(t') \; dt' \right ]
= 0 \;  .
\ee
Therefore the evolution operator has the form
\be
\label{163}
 \hat U(t) = \exp \left \{ -i \int_0^t H(t') \; dt'
\right \} \;  .
\ee
With the total Hamiltonian (\ref{152}), this becomes
\be
\label{164}
 \hat U(t) = \exp \left \{ -i H_A t -i \int_0^t H_M(t') \; dt'
\right \} \;  .
\ee

In view of eigenproblems (\ref{157}) and (\ref{158}), we have
\be
\label{165}
H | nk \rgl = ( E_n + \ep_{nk}(t) ) | nk \rgl \;   .
\ee
Therefore the action of the evolution operator on the vector
$|nk>$ yields
\be
\label{166}
\hat U(t)| nk \rgl = \exp \left \{ -i E_n t -
i \int_0^t \ep_{nk}(t') \; dt' \right \} | nk \rgl  \;   .
\ee

\subsection{Temporal dependence of observables}

Observable quantities are represented by operator averages. Let
us consider the evolution of an observable quantity represented
by an operator $\hat{A}$ acting on $\mathcal{H}_A$. This observable
is given by the average
\be
\label{167}
 \lgl \hat A(t) \rgl \equiv {\rm Tr}_\cH \hat\rho(t) \hat A =
\sum_{mnk} \rho_{mnk}(t) A_{nm} \;  ,
\ee
in which
\be
\label{168}
\rho_{mnk}(t) \equiv \lgl mk | \hat\rho(t) | nk \rgl \; .
\ee
Under the evolution operator (\ref{164}), we have
\be
\label{169}
 \rho_{mnk}(t) = \rho_{mnk}(0) \exp \left \{ - i \om_{mn} t -
i \int_0^t \ep_{mnk}(t') \; dt' \right \} \;  ,
\ee
with
\be
\label{170}
\rho_{mnk}(0) \equiv \lgl mk | \hat\rho(0)| nk \rgl \; ,
\ee
and where the notation is used for the energy differences
\be
\label{171}
\ep_{mnk}(t) \equiv \ep_{mk}(t) - \ep_{nk}(t) \; \qquad
\om_{mn}\equiv E_m - E_n \; .
\ee
Notice that
\be
\label{172}
 \ep_{nnk}(t) = 0 \; .
\ee
Because of the initial condition (\ref{160}), Eq. (\ref{170})
reduces to
\be
\label{173}
 \rho_{mnk}(0) = \rho_{mn}(0) p_k \;  ,
\ee
where
\be
\label{174}
 \rho_{mn}(0) \equiv \lgl m | \hat\rho_A(0)| n \rgl \; ,
\qquad p_k \equiv \lgl k | \hat\rho_D(0) | k \rgl \; .
\ee
The normalization conditions
$$
\sum_n \rho_{nn}(0) = 1 \; , \qquad \sum_k p_k = 1 
$$
are valid. In that way, Eq.(\ref{169}) becomes
\be
\label{175}
 \rho_{mnk}(t) = \rho_{mn}(t) p_k \exp \left \{
- i \int_0^t \ep_{mnk}(t') \; dt' \right \} \; ,
\ee
with
\be
\label{176}
\rho_{mn}(t) = \rho_{mn}(0) \exp ( -i \om_{mn} t) \; .
\ee

Introducing the decoherence factor
\be
\label{177}
D_{mn}(t) \equiv \sum_k p_k \exp \left \{ - i
\int_0^t \ep_{mnk}(t') \; dt' \right \}
\ee
allows us to represent average (\ref{167}) in the form
\be
\label{178}
 \lgl \hat A(t) \rgl = 
\sum_{mn} \rho_{mn}(t) A_{nm} D_{mn}(t) \;  .
\ee

Taking into account properties (\ref{171}) and (\ref{172}), we can
rewrite the above average as
\be
\label{179}
 \lgl \hat A(t) \rgl = \sum_n \rho_{nn} A_{nn} + 
\sum_{m\neq n} \rho_{mn}(t) A_{nm} D_{mn}(t) \;  ,
\ee
with $\rho_{nn} \equiv \rho_{nn}(0)$. This expression is similar
to the average for a quasi-isolated system, but with a different
decoherence factor.

\subsection{Specification of measurement procedure}

To better understand the structure of the decoherence factor
(\ref{177}), we need to specify the action of the measurement
Hamiltonian. Let this Hamiltonian, describing $M$ measurements,
have the form
\be
\label{180}
 H_M(t) = \sum_{j=1}^M \hat X_j f_j(t) \;  ,
\ee
where $f_j(t)$ is a real function and $\hat{X}_j$ is a self-adjoint
operator, not explicitly depending on time. This Hamiltonian 
satisfies the Lyappo-Danilevsky condition (\ref{161}). According to 
Eq. (\ref{158}), the operator $H_M$ and, hence, $\hat{X}_j$, possess 
the eigenvectors $|nk>$. That is, we can write
\be
\label{181}
 \hat X_j| nk \rgl = x_{jnk} | nk \rgl \;  ,
\ee
where $x_{jnk}$ is a real number, characterizing the measurement
impact. As an explicit illustration, we can define $\hat{X}_j$ as
a spectral resolution over projector operators:
$$
\hat X_j = \sum_{nk} x_{jnk} \hat P_{nk} \qquad
\left ( \hat P_{nk} \equiv | nk \rgl \lgl nk | \right ) \;   .
$$

The eigenvalues of $H_M$ acquire the form
\be
\label{182}
\ep_{nk}(t) = \sum_{j=1}^M x_{jnk} f_j(t) \; .
\ee
Respectively, for the transition frequency, defined in Eq. (\ref{171}),
we get
\be
\label{183}
 \ep_{mnk}(t) = \sum_{j=1}^M \Dlt_{jmnk} f_j(t) \;  ,
\ee
where
\be
\label{184}
 \Dlt_{jmnk} \equiv x_{jmk} - x_{jnk} \;  .
\ee
Thence the decoherence factor (\ref{177}) becomes
\be
\label{185}
 D_{mn}(t) = \sum_k p_k \exp \left \{ - i 
\sum_{j=1}^M \Dlt_{jmnk} \vp_j(t) \right \} \;  ,
\ee
with the phase function
\be
\label{186}
 \vp_j(t) \equiv \int_0^t f_j(t') \; dt' \;  .
\ee

Being interested in the long-time behavior of the decoherence
factor, let us assume that the phase function, after the last
measurement, occurring at the time $t_M$, does not depend on
the measurement index $j$,
\be
\label{187}
 \vp_j(t) = \vp(t) \qquad ( t > t_M) \;  .
\ee
Then, after $t > t_M$, the decoherence factor takes the form
\be
\label{188}
 D_{mn}(t) = \sum_k p_k \exp \left \{ - i
\overline\Dlt_{mnk} M \vp(t) \right \} \;  ,
\ee
in which the definition of the mean impact
\be
\label{189}
\overline\Dlt_{mnk} \equiv \frac{1}{M} 
\sum_{j=1}^M \Dlt_{jmnk}
\ee
is used. Finally, introducing the distribution of the measurement
impacts
\be
\label{190}
 p_{mn}(x) \equiv \sum_k p_k \dlt \left ( x - 
\overline\Dlt_{mnk} \right ) \;  ,
\ee
we come to the decoherence factor
\be
\label{191}
 D_{mn}(t) = \int p_{mn}(x) \exp \left \{ - i x M \vp(t) 
\right \} \; dx \;  .
\ee

The measuring device is a macroscopic object, because of which the
spectrum of the measurement Hamiltonian, with respect to the index
$k$ should be treated as continuous. This means that the notation
$\Sigma_k$ has to be understood as integration over $k$.
Consequently, the impact distribution (\ref{190}) can be considered
as a measurable function. By its definition, it is also $L^1$
integrable. Therefore, by the Riemann-Lebesgue lemma,
\be
\label{192}
 D_{mn}(t) \ra 0 \qquad ( M \vp(t) \ra \infty ) \;  .
\ee
That is, nondestructive measurements lead, at large times, to the
system decoherence, provided that either the number of measurements
$M$ or the phase function $\varphi(t)$ increases with time, so that
their product also tends to infinity.

\subsection{Decoherence under nondestructive measurements}

To obtain an explicit form of the decoherence factor, we have to
specify the impact distribution (\ref{190}). Taking for the latter
a Gaussian distribution
\be
\label{193}
p_{mn}(x) = \frac{1}{\sqrt{2\pi} \sgm_{mn} } \;
\exp \left \{ - \; \frac{x^2}{2\sgm_{mn}^2} \right \}
\ee
yields the decoherence factor
\be
\label{194}
 D_{mn}(t) = \exp \left \{ - \; \frac{\sgm_{mn}^2}{2} \;
M^2 \vp^2(t) \right \} \;  .
\ee

In the case of the Lorentz distribution
\be
\label{195}
 p_{mn}(x) = \frac{\sgm_{mn} }{\pi(x^2+\sgm_{mn}^2)} \;  ,
\ee
we have the decoherence factor
\be
\label{196}
 D_{mn}(t) = \exp \{ - \sgm_{mn} M \vp(t) \} \;  .
\ee

Let us consider two opposite types of measurements, instantaneous
and continuous. For {\it instantaneous measurement}, we have
\be
\label{197}
f_j(t) = \dlt(t-t_j) \; , \qquad \vp_j(t) = \Theta(t-t_j) \;  .
\ee
Hence,
\be
\label{198}
\vp(t) \equiv \vp_j(t)|_{t>t_M} = 1 \; .
\ee
And {\it continuous measurement} implies
\be
\label{199}
 f_j(t) = 1 \; , \qquad \vp_j(t) = t \;  .
\ee
In any case, decoherence happens when the number of measurements
$M$ increases [60,135,136].

If measurements are accomplished during all period of observation,
the number of measurements is proportional to the time elapsed.
This can be formalized by the relation
\be
\label{200}
 M = \frac{t}{\Dlt t} \; , \qquad \Dlt t \equiv \frac{1}{M} \;
\sum_{j=1}^M ( t_{j+1} - t_j ) \;  .
\ee

To summarize the results, we omit, for brevity, the indices $m$
and $n$, marking instead the form of the impact distribution
$p(x)$, Gaussian (\ref{193}) or Lorentzian (\ref{195}), and the
type of the measurement, instantaneous or continuous. We assume
relation (\ref{200}) and define the {\it decoherence time}
\be
\label{201}
 t_{dec} \equiv \sqrt{ \frac{\Dlt t}{\sgm} } \;  .
\ee

Then we obtain the decoherence factors, under instantaneous
measurements, for the Gaussian distribution,
\be
\label{202}
 D_{inst}^G(t) = \exp \left \{ -\;
\frac{1}{2} \left ( \frac{t}{t_{dec} } 
\right )^2 \right \} \;  ,
\ee
and for the Lorentzian distribution,
\be
\label{203}
 D_{inst}^L(t) = \exp \left ( -\; \frac{t}{t_{dec} } 
\right ) \;  .
\ee

While in the case of continuous measurements, we find for the
Gaussian distribution,
\be
\label{204}
  D_{cont}^G(t) = \exp \left \{ -\;
\frac{1}{2} \left ( \frac{t}{t_{dec} } 
\right )^4 \right \} \;   ,
\ee
and for the Lorentzian distribution,
\be
\label{205}
 D_{cont}^L(t) = \exp \left \{ -\; \left (\frac{t}{t_{dec} } 
\right )^2 \right \} \;  .
\ee
Continuous measurements result in much faster decoherence than
instantaneous ones.

When decoherence leads to the disappearance of the second term in
the operator average (\ref{179}), this also implies equilibration.
Therefore, equilibration of a finite quantum system can be achieved
by the measurement procedure. As has been emphasized above, the
notion of complete isolation is self-contradictory, since to state
that the system is isolated requires accomplishing measurements
confirming the fact of isolation [60,135,136]. But measurements,
even the nondestructive ones, produce decoherence in the observable
quantities. In the long run, decoherence, induced by measurements,
results in the system equilibration.

\subsection{Existence of time arrow}

The real time in the world, as is known, flows in one direction,
from the past to future. The real time arrow cannot be reversed.
But the microscopic equations of motion are time-symmetric,
allowing one to reverse the direction of time. The problem why
this time symmetry becomes broken, when passing from the
microscopic equations to the observable quantities, is the long
standing puzzle attracting the attention of many. It is not our aim
to give here a discussion of various approaches to this problem.
But we want to stress that this problem finds a simple resolution
in the frame of the equilibration theory of quasi-isolated systems.

The explanation for the existence of the unidirected time arrow
lies in the fact that completely isolated systems do not exist in
nature [139-141]. In the best case, they can be quasi-isolated,
interacting with environment that, even if it does not destroy
and does not essentially disturb the system, anyway, influences
the temporal behavior of observable quantities. Even if the
influence of environment is so weak that it can be neglected,
the considered system is subject to measurement procedures. An
absolutely isolated system is an abstraction that cannot be
observed. As soon as one passes from time-reversible microscopic
equations of motion to the consideration of observables, on has
to take into account that the studied system can only be
quasi-isolated. And the statement that the system is quasi-isolated
necessarily presupposes that this fact is to be confirmed by
measurements. In any case, whether there exists environment acting
on a quasi-isolated system or the latter is subject to at least
nondestructive measurements, there appears decoherence and
equilibration. Each of the latter phenomena influences the time
dependence of observable quantities, making their evolution
irreversible.

The existence for any quasi-isolated system of attenuation, caused
either by environment or by measurements, or by both, leads to equilibration
in the strict sense of limit (\ref{4}), but not merely in the sense of 
time averaged quantities. Even infinitesimally small such an attenuation
principally changes the situation, resulting in the existence of time 
arrow. 

In order to emphasize the importance of even an infinitesimally weak 
nonisolatedness, let us consider two limits: One limit is when the 
measurement dispersion $\sigma$ tends to zero, which corresponds to 
neglecting the influence of measurements or environment. In the case of 
the environment influence, the role of $\sigma$ is played by the 
environment attenuation rate $\gamma$. Another limit is when time tends 
to infinity. For the decoherence factor, in the first case, we have
\be
\label{206}
\lim_{\sgm\ra 0} D(t) = 1 \;   .
\ee
While in the second case,
\be
\label{207}
 \lim_{t\ra \infty} D(t) = 0 \;   .
\ee
These properties lead to the noncommutativity of the limits, as
far as
\be
\label{208}
 \lim_{\sgm\ra 0}\;  \lim_{t\ra \infty} D(t) = 0 \;  ,
\ee
while reversing the limits, we get
\be
\label{209}
\lim_{t\ra \infty} \; \lim_{\sgm\ra 0} D(t) = 1 \;  .
\ee

This noncommutativity of the limits for the decoherence factor
results in the noncommutativity of the limits for the observable
quantities:
\be
\label{210}
\lim_{\sgm\ra 0}\;  \lim_{t\ra \infty} \lgl \hat A(t) \rgl \neq
\lim_{t\ra \infty} \; \lim_{\sgm\ra 0}  \lgl \hat A(t) \rgl \; .
\ee
The noncommutativity of these limits is a pivotal property of 
quasi-isolated systems, exhibiting the existence of time
arrow [139-141].

We may also notice that, if we would like to formally consider negative
time, then the exponentials in the decoherence factors (\ref{139}) and 
(\ref{203}) should be treated as depending on $|t|$. Therefore, in all
cases the average $<\hat{A}(t)>$ equals $<\hat{A}(-t)>$. That is, 
formally the observable quantities are symmetric with respect to time 
inversion, in agreement with microscopic equations of motion. However,
any nonequilibrium statistical state of a quasi-isolated system always 
tends with time to an equilibrium or, at least, to a quasi-equilibrium 
state. This tendency explains the existence of time arrow.

\section{Summary}

Finite quantum systems, because of their important and widespread 
technological applications, have become the topic of intense
investigations, both experimental and theoretical. Nonequilibrium
properties of such systems are of special interest, being less 
understood than equilibrium ones. The principal question is whether
finite quantum systems equilibrate and if so, when and how? 

Cold trapped atoms provide a very convenient laboratory for the 
study of nonequilibrium phenomena. The most interesting experiments
with nonequilibrium trapped atoms are reviewed in the present survey,
and the related numerical simulations are discussed.

Finite quantum systems, strictly speaking, cannot be absolutely 
equilibrium, as far as their statistical states are quasi-periodic. 
But, the studies show that isolated integrable as well as nonintegrable 
quantum systems can equilibrate on average to a quasi-equilibrium state 
with a rather long lifetime, since the recurrence time is usually very 
long. Equilibration to such quasi-stationary states can be faster or 
longer depending on the system integrability. The resulting 
quasi-stationary states are, generally, described by representative 
Gibbs ensembles. In particular cases, these can be microcanonical 
ensembles.    

Nonisolated quantum systems, interacting with a macroscopic bath, 
equilibrate in the strict sense. Equilibration is accompanied by 
decoherence. The time dependence of the decoherence factor is defined by 
the density of states that, in turn, depends on the system parameters. 
The relaxation can be exponential or Gaussian [142].  

Actually, completely isolated finite systems do not exist, but there 
are only quasi-isolated systems. Any finite system is always influenced 
by its environment, maybe weakly, but, anyway, noticeably. Moreover, the 
concept of isolated systems is self-contradictory, since in order to 
state that a system is isolated during a period of time, it is necessary 
to prove this by a series of measurements, whose influence makes the 
system not isolated. Such quasi-isolated systems can equilibrate in the 
strict sense. In some cases, there can exist rare events making the 
resulting state quasi-equilibrium. The latter is equilibrium on average.

In any case, quasi-isolated systems do relax to an either equilibrium or a 
quasi-equilibrium state. This irreversible behavior explains the existence 
of time arrow. 

In this review, the equilibration processes have been described, starting 
from a strongly nonequilibrium state and tending to a steady state. Another
problem of great interest is the opposite process of the development of 
nonequilibrium states from the given equilibrium one. For example, acting 
on trapped Bose atoms by external alternating fields [18,143,144], it is 
possible to generate dynamic transitions from the ground-state Bose-Einstein 
condensate to a vortex superfluid and then to a turbulent superfluid [145-147].
Investigating such transformations from a given equilibrium state to a highly 
nonequilibrium one would allow for the better understanding of the development 
and properties of nonequilibrium states in finite quantum systems.
 
Nonequilibrium effects in finite quantum systems can find numerous 
applications, ranging from various electronic devices to quantum computers.
As has been noticed long tome ago by Schr\"{o}dinger [148] and Bohr [149],  
even the functioning of humans may require, for their correct description, the 
necessity of employing the theory of finite quantum systems. For instance, this 
could be necessary for understanding the functioning of genes and brain. In
turn, understanding the brain activity would allow for the creation of thinking
quantum systems [150].

{\it Acknowledgements}. The author appreciates useful discussions with
V.S. Bagnato. Financial support from the Russian Foundation for Basic
Research is acknowledged.

\end{document}